\documentclass[journal]{IEEEtran}

\usepackage{graphicx}
\usepackage{subfigure}
\usepackage{color}
\usepackage{balance}
\usepackage{graphicx}
\usepackage{enumerate}
\usepackage{subfigure}
\usepackage{algorithm}
\usepackage{algorithmic}
\usepackage{booktabs}
\usepackage{cite}
\usepackage{amssymb}
\usepackage{amsmath}

\usepackage{color}
\usepackage{balance}
\usepackage{graphicx}
\usepackage{enumerate}
\usepackage{subfigure}
\usepackage{algorithm}
\usepackage{algorithmic}
\usepackage{booktabs}
\usepackage{cite}
\usepackage{amssymb}
\usepackage{amsmath}
\usepackage{float}
\usepackage{float}
\usepackage{bm}

\usepackage{fancyhdr}
\usepackage{url}

\begin{document}

\title{Deep Reinforcement Learning Based Intelligent Reflecting Surface for Secure Wireless Communications  }

\author{

  Helin~Yang,~\IEEEmembership{Student Member,~IEEE},
          Zehui~Xiong,~\IEEEmembership{Student Member,~IEEE},        Jun~Zhao,~\IEEEmembership{Member,~IEEE},
   Dusit~Niyato,~\IEEEmembership{Fellow,~IEEE},
  Liang~Xiao,~\IEEEmembership{Senior Member,~IEEE},
  and  Qingqing~Wu,~\IEEEmembership{Member,~IEEE}
        \vspace{-20pt}

\thanks{This research is supported by the National Research Foundation (NRF), Singapore, under Singapore Energy Market Authority (EMA), Energy Resilience, NRF2017EWT-EP003-041, Singapore NRF2015-NRF-ISF001-2277, Singapore NRF National Satellite of Excellence, Design Science and Technology for Secure Critical Infrastructure NSoE DeST-SCI2019-0007, A*STAR-NTU-SUTD Joint Research Grant on Artificial Intelligence for the Future of Manufacturing RGANS1906, Wallenberg AI, Autonomous Systems and Software Program and Nanyang Technological University (WASP/NTU) under grant M4082187 (4080), Singapore Ministry of Education (MOE) Tier 1 (RG16/20), and NTU-WeBank JRI (NWJ-2020-004), Alibaba Group through Alibaba Innovative Research (AIR) Program, Alibaba-NTU Singapore Joint Research Institute (JRI),  Nanyang Technological University (NTU) Startup Grant,  Singapore Ministry of Education Academic Research Fund Tier 1 RG128/18, Tier 1 RG115/19, Tier 1 RT07/19, Tier 1 RT01/19, and Tier 2 MOE2019-T2-1-176,  NTU-WASP Joint Project, Singapore National Research Foundation under its Strategic Capability Research Centres Funding Initiative: Strategic Centre for Research in Privacy-Preserving Technologies \& Systems,  Energy Research Institute @NTU , Singapore NRF National Satellite of Excellence, Design Science and Technology for Secure Critical Infrastructure NSoE DeST-SCI2019-0012, AI Singapore 100 Experiments (100E) programme, NTU Project for Large Vertical Take-Off \& Landing Research Platform, and the Natural Science Foundation of China under Grant 61971366. A part of this paper was accepted in IEEE Global Communications Conference, December 2020, Taipei, Taiwan [40]. \emph{ (Corresponding author: Zehui Xiong.)} }

\thanks{H. Yang,  Z. Xiong, J. Zhao, and D. Niyato are with the School of Computer Science and Engineering, Nanyang Technological University, Singapore 639798 (e-mail: hyang013@e.ntu.edu.sg,  zxiong002@e.ntu.edu.sg, junzhao@ntu.edu.sg, dniyato@ntu.edu.sg).}

\thanks{ L. Xiao  is with the Department of Information and Communication Engineering, Xiamen University, Xiamen 361005, China (e-mail: lxiao@xmu.edu.cn).}

\thanks{Q. Wu  is with State Key Laboratory of Internet of Things for Smart City, University of Macau, Macau, 999078 China. (email:qingqingwu@um.edu.mo).}

}

\maketitle

	 \thispagestyle{fancy}
\pagestyle{fancy}
\lhead{This paper appears in \textbf{IEEE Transactions on Wireless Communications}. \hfill \thepage \\ Please feel free to contact us for questions or remarks. \hfill \url{http://JunZhao.info/} }
\cfoot{}
\renewcommand{\headrulewidth}{0.4pt}
\renewcommand{\footrulewidth}{0pt}

\begin{abstract}

In this paper, we study an intelligent reflecting surface (IRS)-aided wireless secure communication system, where an IRS is deployed to adjust its reflecting elements to secure the communication of multiple legitimate users in the presence of multiple eavesdroppers. Aiming to improve the system secrecy rate, a design problem for jointly optimizing the base station (BS)'s beamforming and the IRS's reflecting beamforming is formulated considering different quality of service (QoS) requirements and time-varying channel conditions. As the system is highly dynamic and complex, and it is challenging to address the non-convex optimization problem, a novel deep reinforcement learning (DRL)-based secure beamforming approach is firstly proposed to achieve the optimal beamforming policy against eavesdroppers in dynamic environments. Furthermore, post-decision state (PDS) and prioritized experience replay (PER) schemes are utilized to enhance the learning efficiency and secrecy performance.  Specifically, a modified PDS  scheme is presented to trace the channel dynamic  and adjust the beamforming policy against channel uncertainty accordingly. Simulation results demonstrate that the proposed deep PDS-PER learning based secure beamforming approach can significantly improve the system secrecy rate and QoS satisfaction probability in IRS-aided secure communication systems. \vspace{5pt}

\textbf{\emph{Index Terms}}---Secure communication, intelligent reflecting surface, beamforming, secrecy rate, deep reinforcement learning.
\end{abstract}

\IEEEpeerreviewmaketitle
\section{Introduction}

\IEEEPARstart{P}{hysical}  layer security (PLS) has attracted increasing attention as an alternative of  cryptography-based techniques for wireless communications [1], where PLS exploits the wireless channel characteristics by using signal processing designs and channel coding to support secure communication services without relying on a shared secret key [1], [2]. So far, a variety of approaches have been reported to improve PLS in wireless communication systems, e.g., cooperative relaying strategies [3], [4], artificial noise-assisted beamforming [5], [6], and cooperative jamming [7], [8]. However, employing a large number of active antennas and relays in PLS systems incurs an excessive hardware cost and the system complexity. Moreover, cooperative jamming and transmitting artificial noise require extra transmit power for security guarantees.

To tackle these shortcomings of the existing approaches [3]-[8], a new paradigm, called intelligent reflecting surface (IRS) [9]-[13], has been proposed as a promising technique to achieve high spectrum efficiency and energy efficiency, and enhance secrecy rate in the fifth generation (5G) and beyond wireless communication systems. In particular, IRS is a uniform planar array which is comprised of a  number of low-cost passive reflecting elements, where each of elements adaptively adjusts its reflection amplitude and/or phase to control the strength and direction of the electromagnetic wave, hence IRS is capable of enhancing and/or weakening the reflected signals at different users [9].  As a result, the reflected signal by IRS can increase the received signal at legitimate users while suppressing the signal at the eavesdroppers [9]-[13]. Hence, from the PLS perspective, some innovative studies have been recently devoted to performance optimization for IRS-aided secure communications [14]-[25].

 \subsection{Related Works}

Initial studies on IRS-aided secure communication systems have reported in [14]-[17], where a simple system model with only a single-antenna legitimate user and a single-antenna eavesdropper was considered in these works. The authors in [14] and [15] applied the alternative optimization (AO) algorithm to jointly optimize the transmit beamforming vector at the base station (BS)  and the  phase elements at the IRS for the maximization of the secrecy rate, but they did not extend their models to multi-user IRS-assisted secure communication systems. To minimize the transmit power at the BS subject to the secrecy rate constraint, the authors in [18] utilized AO solution and semidefinite programming (SDP) relaxation to address the optimization problem with the objective to jointly optimize the power allocation and the IRS reflecting beamforming. In addition, Feng $et~al$. [19] also studied the secure transmission framework with an IRS to minimize the system transmit power in cases of rank-one and full-rank BS-IRS links, and derived a closed-form expression of beamforming matrix. Different from these studies [14]-[19] which considered only a single eavesdropper, secure communication systems comprising multiple eavesdroppers were investigated in [20]-[22]. Chen $et~al$. [20] presented a minimum-secrecy-rate maximization design to provide secure communication services for multiple legitimate users while keeping them secret from multiple eavesdroppers in an IRS-aided multi-user multiple-input single-output (MISO) system, but the simplification of the optimization problem may cause a performance loss. The authors in [23] and [24] studied an IRS-aided multiple-input multiple-output (MIMO) channel, where a multi-antenna BS transmits data stream to a multi-antenna legitimate user in the presence of an eavesdropper configured with multiple antennas, and a suboptimal secrecy rate maximization approach was presented to optimize the beamforming policy. In addition to the use of AO or SDP in the system performance optimization, the minorization-maximization (MM) algorithm was recently utilized to optimize the joint transmit beamforming at the BS and phase shift coefficient at the IRS [16], [23].

Moreover, the authors in [22] and [25] employed the artificial noise-aided beamforming for IRS-aided MISO secure communication systems to improve the system secrecy rate, and an AO based solution was applied to jointly optimize the BS's beamforming, artificial noise interference vector and IRS's reflecting beamforming with the goal to maximize the secrecy rate. All these existing studies [14]-[20], [22]-[25] assumed that perfect channel state information (CSI) of legitimate users or eavesdroppers is available at the BS, which is not a practical assumption. The reason is that acquiring perfect CSI at the BS is challenging since the corresponding CSI may be outdated when the channel is time-varying due to the transmission delay, processing delay, and high mobility of users.  Hence, Yu $et~al$. [21] investigated an optimization problem with considering the impacts of outdated CSI of the eavesdropping channels in an IRS-aided secure communication system, and a robust algorithm was proposed to address the optimization problem in the presence of multiple eavesdroppers.

The above mentioned studies [14]-[25] mainly applied the traditional optimization techniques e.g., AO, SDP or MM algorithms to jointly optimize the BS's beamforming and the IRS's reflecting beamforming in IRS-aided secure communication systems, which are less efficient for large-scale systems. Inspired by the recent advances of artificial intelligence (AI), several works attempted to utilize  AI algorithms to optimize IRS's reflecting beamforming [26]-[29]. Deep learning (DL) was exploited to search the optimal IRS reflection matrices that maximize the achievable system rate in an IRS-aided communication system, and the simulation demonstrated that DL significantly outperforms conventional algorithms. Moreover, the authors in  [28] and [29] proposed deep reinforcement learning (DRL) based approach to address the non-convex optimization problem, and the phase shifts at the IRS are optimized effectively. However, the works [26]-[29] merely considered to maximize the system achievable rate of a single user without considering the scenario of multiple users, secure communication and imperfect CSI in their models. The authors in [30] and [31] applied reinforcement learning (RL) to achieve  smart beamforming at the BS against an eavesdropper in complex environments, but the IRS-aided secure communication system needs to optimize the IRS's reflect beamforming in addition to the BS's transmit beamforming. To the best of our knowledge, RL or DRL has not been explored yet in prior works to optimize both the BS's transmit beamforming and the IRS's reflect beamforming in dynamic IRS-aided secure communication systems, under the condition of multiple eavesdroppers and imperfect CSI, which thus motivates this work.

\subsection{Contributions}

In this paper, we investigate an IRS-aided secure communication system with the objective to maximize the system secrecy rate of multiple legitimate users in the presence of multiple eavesdroppers under realistic time-varying channels, while guaranteeing quality of service (QoS) requirements of legitimate users. A novel DRL-based secure beamforming approach is firstly proposed to jointly optimize the beamforming matrix at the BS and the reflecting beamforming matrix (reflection phases) at the IRS in dynamic environments. The major contributions of this paper are summarized as follows:
\begin{itemize}
\item  The physical secure communication based on IRS with multiple eavesdroppers is investigated under the condition of time-varying channel coefficients in this paper. In addition, we formulate a joint BS's transmit beamforming and IRS's reflect beamforming optimization problem with the goal  of maximizing the system secrecy rate while considering the QoS requirements of legitimate users.
\item  An RL-based intelligent beamforming framework is presented to achieve the optimal BS's beamforming and the IRS's reflecting beamforming, where the central controller intelligently optimizes the beamforming policy by using a Markov decision process (MDP) according to the instantaneous observations from dynamic environment. Specifically, a QoS-aware reward function is constructed by covering both the secrecy rate and users' QoS requirements into the learning process.
\item  A DRL-based secure beamforming approach is proposed to improve the learning efficiency and secrecy performance by fully exploiting the information of complex structure of the beamforming policy domain,  where a modified post-decision state (PDS) learning is presented to trace the channel dynamic against channel uncertainty, and prioritized experience replay (PER) is applied to enhance the learning efficiency.
\item  Extensive simulation results are provided to demonstrate the effectiveness of the proposed deep PDS-PER leaning based secure beamforming approach in terms of improving the secrecy rate and the QoS satisfaction probability, compared with other existing approaches. For instance, the proposed learning approach achieves the secrecy rate and QoS satisfaction level improvements of 17.21\% and 8.67\%, compared with the approach [14] in  time-varying channel condition.
\end{itemize}

The rest of this paper is organized as follows. Section II presents the system model and problem formulation. The optimization problem is formulated as an RL problem in Section III. Section IV proposes a deep PDS-PER based secure beamforming approach. Section V provides simulation results and Section VI concludes the paper.

Notations: In this paper, vectors and matrices are represented by
Boldface lowercase and uppercase letters, respectively.
${\rm{Tr}}( \cdot )$, ${( \cdot )^ * }$ and ${( \cdot )^H}$
denote the trace, the conjugate and the conjugate transpose
operations, respectively. $| \cdot |$ and  $|| \cdot ||$ stand for
the absolute value of a scalar and the Euclidean norm of a vector
or matrix, respectively. $\mathbb{E}[ \cdot ]$ denotes the
expectation operation. ${\mathbb{C}^{M \times N}}$ represents the
space of complex-valued matrices.

\section{System Model and Problem Formulation}

\subsection{System Model}

\begin{figure}
\centering
\includegraphics[width=0.82\columnwidth]{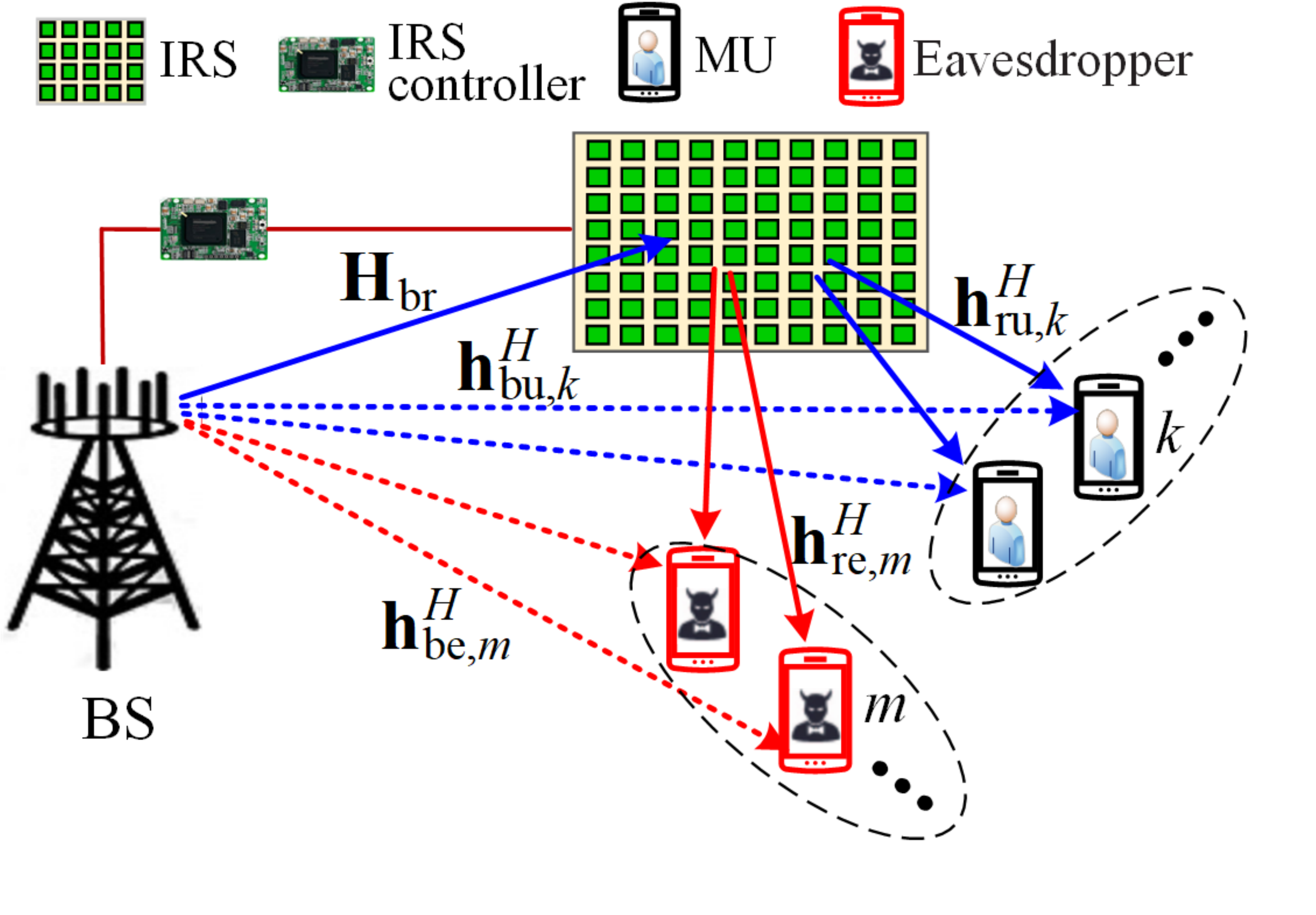}
\caption{{IRS-aided secure communication under multiple
eavesdroppers.} } \label{fig:Schematic}
\end{figure}

  We consider an IRS-aided secure communication
system, as shown in Fig. 1, where the BS is equipped with $N$
antennas to serve $K$ single-antenna legitimate mobile users (MUs)
in the presence of $M$ single-antenna eavesdroppers. An IRS with $L$
reflecting elements is deployed in the system to assist secure
wireless communications from the BS to the MUs. The IRS is
equipped with a controller to  coordinate with the BS. For the ease of
practical implementation, the maximal reflection without power loss at the IRS
is considered since the reflecting elements are designed to
maximize the reflected desired signal power to the MUs [13]-[23].
In addition, unauthorized eavesdroppers aim to eavesdrop any of
the data streams of the MUs.  Hence, the use of
reflecting beamforming at IRS is also investigated to improve the
achievable secrecy rate at the MUs while suppressing the
wiretapped data rate at the eavesdroppers. In addition, we explicitly state that the eavesdroppers cannot collide [5], [6], [18]-[21].

 Let  $\mathcal{K} = \{
1,2,\ldots,K\} $,   $\mathcal{M} = \{ 1,2,\ldots,M\} $ and $\mathcal{L}
= \{ 1,2,\ldots,L\} $ denote the MU set, the eavesdropper set and the
IRS reflecting element set, respectively. Let  ${\bf{H}_{{\rm{br}}}}
\in {\mathbb{C}^{L \times N}}$,  ${\bf{h}}_{{\rm{bu,}}k}^H \in
{\mathbb{C}^{1 \times N}}$, ${\bf{h}}_{{\rm{ru,}}k}^H \in {\mathbb{C}^{1
\times L}}$, ${\bf{h}}_{{\rm{be,}}m}^H \in {\mathbb{C}^{1 \times N}}$,
and ${\bf{h}}_{{\rm{re,}}m}^H \in {\mathbb{C}^{1 \times L}}$ denote the
channel coefficients from the BS to the IRS, from the BS to the
$k$-th MU, from the IRS to the $k$-th MU, from the BS to the
$m$-th eavesdropper, and from the IRS to the $m$-th eavesdropper,
respectively. All the above mentioned channel coefficients in the
system are assumed to be small-scale fading with path loss  which follows the
Rayleigh fading model [11]-[14], [21]. Let ${\bf{\Psi }} =
{\rm{diag}}({\chi _1}{e^{j{\theta _1}}},{\chi _2}{e^{j{\theta
_2}}},\ldots,{\chi _L}{e^{j{\theta _L}}})$ denote the reflection
coefficient matrix associated with effective phase shifts at the
IRS, where  ${\chi _l} \in [0,1]$  and ${\theta _l} \in [0,2\pi ]$
denote the amplitude reflection factor and the phase shift
coefficient on the combined transmitted signal, respectively.
As each phase shift is desired  to be designed to achieve full reflection, we consider that ${\chi
_l} = 1$, $\forall l \in \mathcal{L}$ in the sequel of the paper.

At the BS side, the beamforming vector for the $k$-th MU is
denoted as  ${{\bf{v}}_k} \in {\mathbb{C}^{N \times 1}}$, which is
the continuous linear precoding [11]-[16], [23]. Thus, the
transmitted signal for all MUs at the BS is written as  ${\bf{x}} =
\sum\nolimits_{k = 1}^K {{{\bf{v}}_k}{s_k}} $, where ${s_k}$ is
the transmitted symbol for the $k$-th MU  which can be  modelled as independent
and identically distributed (i.i.d.) random variables with zero mean and
unit variance [11]-[16], [23], and ${s_k} \sim \mathcal{CN}(0,1)$. The total transmit power at the BS
is subject to the maximum power constraint:
\begin{equation}
\begin{split}
\mathbb{E}[||{\bf{x}}{||^2}] = {\rm{Tr(}}{\bf{V}}{\bf{V}}^{H}{\rm{)}} \le {P_{\max }}
\end{split}
\end{equation}
where  ${\bf{V}} \buildrel \Delta \over =
[{{\bf{v}}_1},{{\bf{v}}_2},\ldots,{{\bf{v}}_K}] \in {\mathbb{C}^{M
\times K}}$, and ${P_{\max }}$ is the maximum transmit power
at the BS.

When the BS transmits a secret message to the $k$-th MU, the MU
will receive the signal from the BS and the reflected signal from
the IRS.
Accordingly, the received signal at MU $k$ can be given by
\begin{equation}
\begin{split}
\begin{array}{l}
{y_k} = \underbrace {\left( {{\bf{h}}_{{\rm{ru,}}k}^H{\bf{\Psi }}{{\bf{H}}_{{\rm{br}}}} + {\bf{h}}_{{\rm{bu,}}k}^H} \right){{\bf{v}}_k}{s_k}}_{{\rm{desired}}\;{\rm{signal}}}\\
\;\;\;\; + \underbrace {\sum\limits_{i \in {\mathcal{K}},i \ne k} {\left( {{\bf{h}}_{{\rm{ru,}}k}^H{\bf{\Psi }}{{\bf{H}}_{{\rm{br}}}} + {\bf{h}}_{{\rm{bu,}}k}^H} \right){{\bf{v}}_i}{s_i}} }_{{\rm{inter - user interference}}} + {n_k}
\end{array}
\end{split}
\end{equation}

where ${n_k}$  denotes the additive complex Gaussian noise (AWGN)
with the with zero mean and variance  $\delta _k^2$ at the $k$-th
MU. In (2), we observe that in addition to the received desired
signal, each MU also suffers inter-user interference (IUI) in the
system. In addition, the received signal at eavesdropper $m$ is
expressed by
\begin{equation}
\begin{split}
{y_m} = \left( {{\bf{h}}_{{\rm{re,}}m}^H{\bf{\Psi }}{{\bf{H}}_{{\rm{br}}}} + {\bf{h}}_{{\rm{be,}}m}^H} \right)\sum\limits_{k \in {\mathcal{K}}} {{{\bf{v}}_k}{s_k}}  + {n_m}
\end{split}
\end{equation}
where ${n_m}$  is the AWGN of eavesdropper $m$ with the variance
$\delta _m^2$ .

{\color{blue}}

In practical systems, it is not easy  for the BS and the IRS to
obtain perfect CSI [9], [21]. This is  due to the fact that both the
transmission delay and processing delay exist, as well as the
mobility of the users. Therefore, CSI is outdated at the time
when the BS and the IRS transmit the data stream to MUs [21].
Once this outdated CSI is employed for beamforming, it will lead
to a negative effect on the demodulation at the MUs, thereby
leading to substantial performance loss [21]. Therefore, it is necessary to
consider outdated CSI in the IRS-aided secure communication
system.

Let ${T_{{\rm{delay}}}}$ denote the delay between the outdated CSI
and the real-time CSI. In other words, when the BS receives the
pilot sequences sent from the MUs at the time slot $t$, it will
complete the channel estimation process and begin to transmit data
stream to the MUs at the time slot $t + {T_{{\rm{delay}}}}$.
Hence, the relation between the outdated channel vector ${\bf{h}}(t)$ and the
real-time channel vector ${\bf{h}}(t +
{T_{{\rm{delay}}}})$ can be expressed by
\begin{equation}
\begin{split}
{\bf{h}}(t + {T_{{\rm{delay}}}}) = \rho {\bf{h}}(t) + \sqrt {1 - {\rho ^2}}
\hat {\bf{h}}(t + {T_{{\rm{delay}}}}).
\end{split}
\end{equation}

In (4), $\hat {\bf{h}}(t + {T_{{\rm{delay}}}})$  is independent
identically distributed with ${\bf{h}}(t)$ and  ${\bf{h}}(t +
{T_{{\rm{delay}}}})$, and it is with zero-mean and unit-variance
complex Gaussian entries. $\rho $ is the autocorrelation function
(outdated CSI coefficient) of the channel gain ${\bf{h}}(t)$ and  $0 \le
\rho  \le 1$, which is given by
\begin{equation}
\begin{split}
\rho  = {J_0}(2{\pi _{{\rm{pi}}}}{f_D}{T_{{\rm{delay}}}})
\end{split}
\end{equation}
where  ${J_0}( \cdot )$ is the zeroth-order Bessel function of the
first kind, ${f_D}$ is the Doppler spread which is generally a
function of the velocity ($\upsilon $) of the transceivers, the
carrier frequency (${f_c}$) and the speed of light $(c)$, i.e.,
${f_D} = \upsilon {f_c}/c$. Note that $\rho  = 1$ indicates the
outdated CSI effect is eliminated, whereas $\rho  = 0$ represents
no CSI.

As the outdated CSI introduces the channel uncertainty in practical dynamic systems, the actual channel coefficients can be rewritten as
\begin{equation}
\begin{split}
\begin{array}{l}
{{\bf{h}}_{{\rm{bu}},k}} = {{{\bf{\tilde h}}}_{{\rm{bu}},k}} + \Delta {{\bf{h}}_{{\rm{bu}},k}},\;\forall k \in \mathcal{K},\\
{{\bf{h}}_{{\rm{ru}},k}} = {{{\bf{\tilde h}}}_{{\rm{ru}},k}} + \Delta {{\bf{h}}_{{\rm{ru}},k}},\;\forall k \in \mathcal{K},\\
{{\bf{h}}_{{\rm{be}},m}} = {{{\bf{\tilde h}}}_{{\rm{be}},m}} + \Delta {{\bf{h}}_{{\rm{be}},m}},\;\forall m \in \mathcal{M},\\
{{\bf{h}}_{{\rm{re}},m}} = {{{\bf{\tilde h}}}_{{\rm{re}},m}} + \Delta {{\bf{h}}_{{\rm{re}},m}},\;\forall m \in \mathcal{M},
\end{array}
\end{split}
\end{equation}
where  ${{\bf{\tilde h}}_{{\rm{bu}},k}}$,  ${{\bf{\tilde h}}_{{\rm{ru}},k}}$,  ${{\bf{\tilde h}}_{{\rm{be}},m}}$ and ${{\bf{\tilde h}}_{{\rm{re}},m}}$ denote the estimated channel vectors;  $\Delta {{\bf{h}}_{{\rm{bu}},k}}$, $\Delta {{\bf{h}}_{{\rm{ru}},k}}$, $\Delta {{\bf{h}}_{{\rm{be}},m}}$ and $\Delta {{\bf{h}}_{{\rm{re}},m}}$ are the corresponding channel error vectors. In the paper, generally, the channel error vectors of each MU and each eavesdropper can be bounded with respect to the Euclidean norm by using norm-bounded error model, i.e.,
\begin{equation}
\begin{split}
\begin{array}{l}
||\Delta {{\bf{h}}_{{\rm{bu}}}}|{|^2} \le {({\varsigma _{{\rm{bu}}}})^2},\;\;||\Delta {{\bf{h}}_{{\rm{ru}}}}|{|^2} \le {({\varsigma _{{\rm{ru}}}})^2},\\
||\Delta {{\bf{h}}_{{\rm{be}}}}|{|^2} \le {({\varsigma _{{\rm{be}}}})^2},\;\;||\Delta {{\bf{h}}_{{\rm{re}}}}|{|^2} \le {({\varsigma _{{\rm{re}}}})^2},
\end{array}
\end{split}
\end{equation}
where  ${\varsigma _{{\rm{bu}}}}$, ${\varsigma _{{\rm{ru}}}}$, ${\varsigma _{{\rm{be}}}}$, and  ${\varsigma _{{\rm{re}}}}$ refer to the radii of the deterministically bounded error regions.

Under the channel uncertainty model, the achievable rate of the $k$-th MU is given by

\begin{equation}
\begin{split}
R_k^{\rm{u}} = {\log _2}\left( {1 + \frac{{{{\left|
{({\bf{h}}_{{\rm{ru,}}k}^H{\bf{\Psi }}{{\bf{H}}_{{\rm{br}}}} +
{\bf{h}}_{{\rm{bu,}}k}^H){{\bf{v}}_k}} \right|}^2}}}{{{{|
{\sum\limits_{i \in {\mathcal{K}},i \ne k} {({\bf{h}}_{{\rm{ru,}}k}^H{\bf{\Psi
}}{{\bf{H}}_{{\rm{br}}}} + {\bf{h}}_{{\rm{bu,}}k}^H){{\bf{v}}_i}} } |}^2}
+ \delta _k^2}}} \right).
\end{split}
\end{equation}

If the $m$-th eavesdropper attempts to eavesdrop the signal of the $k$-th MU, its achievable rate can be expressed by
\begin{equation}
\begin{split}
\begin{array}{l}
R_{m,k}^{\rm{e}} = \\
{\log _2}\left( {1 + \frac{{{{\left|
{({\bf{h}}_{{\rm{re,}}m}^H{\bf{\Psi }}{{\bf{H}}_{{\rm{br}}}} +
{\bf{h}}_{{\rm{be,}}m}^H){{\bf{v}}_k}} \right|}^2}}}{{{{|
{\sum\limits_{i \in {\mathcal{K}}, i \ne k} {({\bf{h}}_{{\rm{re,}}m}^H{\bf{\Psi
}}{{\bf{H}}_{{\rm{br}}}} + {\bf{h}}_{{\rm{be,}}m}^H){{\bf{v}}_i}} } |}^2}
+ \delta _m^2}}} \right).
\end{array}
\end{split}
\end{equation}

Since each eavesdropper can eavesdrop any of the $K$ MUs' signal,
according to [14]-[25], the achievable individual secrecy
rate from the BS to the $k$-th MU can be expressed by
\begin{equation}
\begin{split}
R_k^{\sec } = {\left[ {R_k^{\rm{u}} - \mathop {\max
}\limits_{\forall m} R_{m,k}^{\rm{e}}} \right]^ + }
\end{split}
\end{equation}
where  ${[z]^ + } = \max (0,z)$.

\subsection{Problem Formulation}

 Our objective is to jointly optimize the robust BS's transmit beamforming matrix ${\bf{V}}$ and the robust IRS's reflecting beamforming matrix  ${\bf{\Psi }}$  from the system beamforming codebook $\mathcal{F}$ to maximize the worst-case secrecy rate with the worst-case secrecy rate and data rate constraints, the total BS transmit power constraint and the  IRS  reflecting unit constraint. As
such, the optimization problem is formulated
as
\begin{equation}
\begin{split}
\begin{array}{l}
\mathop {\max }\limits_{{\bf{V}},{\bf{\Psi }}} \mathop {\min }\limits_{\{ \Delta {\bf{h}}\} } \sum\limits_{k \in \mathcal{K}} {R_k^{\sec }} \\
s.t.\;\;({\rm{a}}):\;R_k^{\sec } \ge R_k^{\sec ,\min },\;\forall k \in \mathcal{K},\\
\;\;\;\;\;\;\;({\rm{b}}):\;\mathop {\min }\limits_{\scriptstyle||\Delta {{\bf{h}}_{{\rm{bu}}}}|{|^2} \le {({\varsigma _{{\rm{bu}}}})^2},\hfill\atop
\scriptstyle||\Delta {{\bf{h}}_{{\rm{ru}}}}|{|^2} \le {({\varsigma _{{\rm{ru}}}})^2}\hfill} (R_k^{\rm{u}}) \ge R_k^{\min },\;\forall k \in \mathcal{K},\\
\;\;\;\;\;\;\;({\rm{c}}):\;{\rm{Tr}}\left( {{\bf{V}}{{\bf{V}}^H}} \right) \le {P_{\max }},\\
\;\;\;\;\;\;\;({\rm{d}}):\;|\chi {e^{j{\theta _l}}}| = 1,\;0 \le {\theta _l} \le 2\pi ,\;\forall l \in \mathcal{L},
\end{array}
\end{split}
\end{equation}
where $R_k^{\sec {\rm{,min}}}$ is the target secrecy rate of the
$k$-th MU, and $R_k^{{\rm{min}}}$ denotes its target data rate.
The constraints in  (11a) and (11b) are imposed to satisfy the worst-case secrecy rate and data rate requirements,
respectively. The constraint in  (11c) is set to satisfy the BS's maximum power
constraint.  The constraint in  (11d) is the constraint of the
IRS reflecting elements.
Obviously, it is challenging to obtain an optimal solution to
the optimization (11), since the objective function in (11) is
non-concave with respect to either  ${\bf{V}}$ or  ${\bf{\Psi }}$,
and the coupling of the optimization variables (${\bf{V}}$ and
${\bf{\Psi }}$) and the unit-norm constraints in (11d) are
non-convex. In addition,  we would consider the robust beamforming design to maximize the worst-case achievable secrecy rate of the system while guaranteeing the worst-case constraints.

\section{Problem Transformation Based on RL}

The optimization problem given in (11) is difficult to address
as it is a non-convex problem. In
addition, in realistic IRS-aided secure communication systems, the
capabilities of MUs, the channel quality, and the service
applications will change dynamically.  Moreover, the problem in
(11) is just a single time slot optimization problem, which may
converge to a suboptimal solution and obtain the greedy-search
like performance due to the ignorance of the historical system
state and the long term benefit. Hence, it is generally infeasible
to apply the traditional optimization techniques (AO, SDP, and MM)
to achieve an effective secure beamforming policy in uncertain
dynamic environments.

Model-free RL is a dynamic programming tool which can be adopted
to solve the decision-making problem by learning the optimal
solution in dynamic environments [32]. Hence, we model the secure
beamforming optimization problem as an RL problem. In RL, the
IRS-aided secure communication system is treated as an
environment, the central controller at the BS
is regarded as a learning agent. The key elements of RL are
defined as follows.

\textbf{State space:}  Let ${\mathcal{S}}$ denote the
system state space. The current system state  $s \in
{\mathcal{S}}$ includes the channel information of all users, the secrecy rate, the transmission data rate of the last
time slot and the QoS satisfaction level, which is defined as
\begin{equation}
\begin{split}
\begin{array}{l}
s = \left\{ {{{\{ {{\bf{h}}_k}{\rm{\} }}}_{k \in{\mathcal{K}}}},{{\{ {{\bf{h}}_m}{\rm{\} }}}_{m \in {\mathcal{M}}}},{{{\rm{\{ }}R_k^{\sec }{\rm{\} }}}_{k \in {\mathcal{K}}}},{{\{ {R_k}\} }_{k \in {\mathcal{K}}}},} \right.\\
\;\;\;\;\;\;\left. {{{\{ {\rm{Qo}}{{\rm{S}}_k}\} }_{k \in {\mathcal{K}}}}} \right\}
\end{array}
\end{split}
\end{equation}

where ${{\bf{h}}_k}$ and ${{\bf{h}}_m}$ are  the channel
coefficients of the $k$-th MU and $m$-th eavesdropper,
respectively. ${\rm{Qo}}{{\rm{S}}_k}$ is the feedback QoS satisfaction level of the $k$-th
MU, where the QoS satisfaction level consists of  both the minimum secrecy rate satisfaction level in (11a) and the minimum data rate satisfaction level in (11b). Other parameters in (12) are already defined in Section II.

\textbf{Action space:}  Let ${\mathcal{A}}$ denote the
system action space. According to the observed system state $s$,
the central controller chooses the beamforming vector ${\{
{{\bf{v}}_k}\} _{k \in {\mathcal{K}}}}$ at the BS and the IRS
reflecting beamforming coefficient (phase shift)  ${\{ {\theta
_l}\} _{l \in {\mathcal{L}}}}$  at the IRS. Hence, the action $a
\in {\mathcal{A}}$ can be defined by
\begin{equation}
\begin{split}
a = \left\{ {{{\{ {{\bf{v}}_k}\} }_{k \in {\mathcal{K}}}},{{\{
{\theta _l}\} }_{l \in {\mathcal{L}}}}} \right\}.
\end{split}
\end{equation}

\textbf{Transition probability:}  Let ${\mathcal{T}}(s'|s,a)$
represent the transition probability, which is the probability of
transitioning to a new state  $s' \in {\mathcal{S}}$, given the
action  $a$ executed in the sate $s$.

\textbf{Reward function:} In RL, the reward acts as a signal
to evaluate how good the secure beamforming policy is when the
agent executes an action at a current state. The system
performance will be enhanced when the reward function at each
learning step correlates with the desired objective. Thus, it is
important to design an efficient reward function to improve the
MUs' QoS satisfaction levels.

In this paper, the reward function represents the optimization
objective, and our objective is to maximize the system secrecy
rate of all MUs while guaranteeing their QoS requirements. Thus,
the presented QoS-aware reward function is expressed as
\begin{equation}
\begin{split}
r = \underbrace {\sum\limits_{k \in {\mathcal{K}}} {R_k^{\sec }}
}_{{\rm{part}}\;{\rm{1}}} - \underbrace {\sum\limits_{k \in
{\mathcal{K}}} {{\mu _1}p_k^{\sec }} }_{{\rm{part}}\;2} -
\underbrace {\sum\limits_{k \in {\mathcal{K}}} {{\mu
_2}p_k^{\rm{u}}} }_{{\rm{part}}\;3}
\end{split}
\end{equation}
where
\begin{equation}
\begin{split}
p_k^{\sec } = \left\{ \begin{array}{l}
1,\;{\rm{if}}\;R_k^{\sec } < R_k^{\sec {\rm{,min}}},\forall k \in {\mathcal{K}},\\
0,\;{\rm{otherwise}}{\rm{,}}
\end{array} \right.
\end{split}
\end{equation}

\begin{equation}
\begin{split}
p_k^{\rm{u}} = \left\{ \begin{array}{l}
1,\;{\rm{if}}\;{R_k} < R_k^{{\rm{min}}},\forall k \in {\mathcal{K}},\\
0,\;{\rm{otherwise}}{\rm{.}}
\end{array} \right.
\end{split}
\end{equation}

 In (14), the part 1 represents the immediate utility (system
secrecy rate), the part 2 and the part 3 are the cost functions
which are defined as the unsatisfied secrecy rate requirement and
the unsatisfied minimum rate requirement, respectively. The
coefficients ${\mu _1}$ and  ${\mu _2}$ are the positive constants
of the part 2 and the part 3 in (14) , respectively, and they are used to balance
the utility and cost [33]-[35].

The goals of (15) and (16) are to impose the QoS satisfaction
levels of both the secrecy rate and the minimum data rate
requirements, respectively. If the QoS requirement is satisfied in
the current time slot, then   $p_k^{\sec } = {\rm{0}}$ or
$p_k^{\rm{u}} = {\rm{0}}$, indicating that there is no
punishment of the reward function due to the successful QoS
guarantees.

The goal of the learning agent is to search for an optimal policy
${\pi ^ * }$ ($\pi $  is a mapping from states in ${\mathcal{S}}$
to the probabilities of choosing an action in ${\mathcal{A}}$:
$\pi (s):{\mathcal{S}} \to {\mathcal{A}}$) that maximizes the
long-term expected discounted reward, and the cumulative
discounted reward function can be defined as
\begin{equation}
\begin{split}
{U_t} = \sum\limits_{\tau  = 0}^\infty  {{\gamma _\tau }{r_{t +
\tau  + 1}}}
\end{split}
\end{equation}
where $\gamma  \in (0,1]$ denotes the discount factor. Under a
certain policy  $\pi $, the state-action function of the agent
with a state-action pair ($s$, $a$) is given by
\begin{equation}
\begin{split}
{Q^\pi }({s_t},{a_t}) = {\mathbb{E}_\pi }\left[ {{U_t}|{s_t} =
s,{a_t} = a} \right].
\end{split}
\end{equation}

The conventional Q-Learning algorithm can be adopted to learn the
optimal policy. The key objective of Q-Learning is to update
Q-table by using the Bellman's equation as follows:
\begin{equation}
\begin{split}
\begin{array}{l}
{Q^\pi }({s_t},{a_t}) = {\mathbb{E}_\pi }\left[ {{r_t} + \gamma \sum\limits_{{s_{t + 1}} \in {\mathcal{S}}} {{\mathcal{T}}({s_{t + 1}}|{s_t},{a_t})} } \right.\\
\;\;\;\;\;\;\;\;\;\;\left. {\sum\limits_{{a_{t + 1}} \in {\mathcal{A}}} {\pi ({s_{t + 1}},{a_{t + 1}}){Q^\pi }({s_{t + 1}},{a_{t + 1}})} } \right]
\end{array}
\end{split}
\end{equation}

The optimal action-value function in (17) is equivalent to the
Bellman optimality equation, which is expressed by
\begin{equation}
\begin{split}
{Q^ * }({s_t},{a_t}) = {r_t} + \gamma \mathop {\max
}\limits_{{a_{t + 1}}} {Q^*}({s_{t + 1}},{a_{t + 1}})
\end{split}
\end{equation}
and the state-value function is achieved as follows:
\begin{equation}
\begin{split}
V({s_t}) = \mathop {\max }\limits_{{a_t} \in {\mathcal{A}}}
Q({s_t},{a_t}).
\end{split}
\end{equation}

In addition, the Q-value is updated as follows:
\begin{equation}
\begin{split}
\begin{array}{l}
{Q_{t + 1}}({s_t},{a_t}) = (1 - {\alpha _t}){Q_t}({s_t},{a_t}) +
{\alpha _t}\left( {{r_t} + \gamma {V_t}({s_{t{\rm{ + 1}}}})}
\right)
\end{array}
\end{split}
\end{equation}
where  ${\alpha _t} \in (0,1]$ is the learning rate. Q-Learning
generally constructs a lookup Q-table $Q(s,a)$, and the agent
selects actions based on the greedy policy for each learning step
[32]. In the $\varepsilon  - $greedy policy, the agent chooses the
action with the maximum Q-table value with probability ${\rm{1}} -
\varepsilon $, whereas a random action is picked with probability
$\varepsilon $ to avoid achieving stuck at non-optimal policies
[32]. Once the optimal Q-function ${Q^ * }(s,a)$  is achieved, the
optimal policy is determined by
\begin{equation}
\begin{split}
{\pi ^ * }(s,a) = \arg \mathop {\max }\limits_{a \in
{\mathcal{A}}} {Q^ * }(s,a).
\end{split}
\end{equation}

\section{Deep PDS-PER Learning Based Secure Beamforming}

The secure beamforming policy discussed in Section III can be
numerically achieved by using Q-Learning, policy gradient, and
deep Q-Network (DQN) algorithms [32]. However, Q-Learning is not
an efficient learning algorithm because it cannot deal with
continuous state space and it has slow learning convergence speed.
The policy gradient algorithm has the ability to handle continuous
state-action spaces, but it may converge to a suboptimal
solution. In addition, it is intractable for Q-learning and policy
gradient algorithms to solve the optimization problem under
high-dimensional input state space. Although DQN performs well in
policy learning  under  high-dimensional state
space, its non-linear Q-function estimator may lead to unstable learning process.

Considering the fact that the IRS-aided secure communication
system has high-dimensional and high-dynamical characteristics
according to the system state that is defined in (12) and
uncertain CSI that is shown in (4) and (6), we propose a deep PDS-PER
learning based secure beamforming approach, as shown in Fig. 2,
where PDS-learning and PER mechanisms are utilized to enable the
learning agent to learn and adapt faster in dynamic environments.
In detail, the agent utilizes the observed state (i.e, CSI,
previous secrecy rate, QoS satisfaction level), the feedback
reward from environment as well as the historical experience from
the replay buffer to train its learning model. After that, the
agent employs the trained model to make decision (beamforming
matrices ${\bf{V}}$ and  ${\bf{\Psi }}$) based on its learned
policy. The  procedures of the proposed learning based secure
beamforming are provided in the following subsections.

\begin{figure}
\centering
\includegraphics[width=0.925\columnwidth]{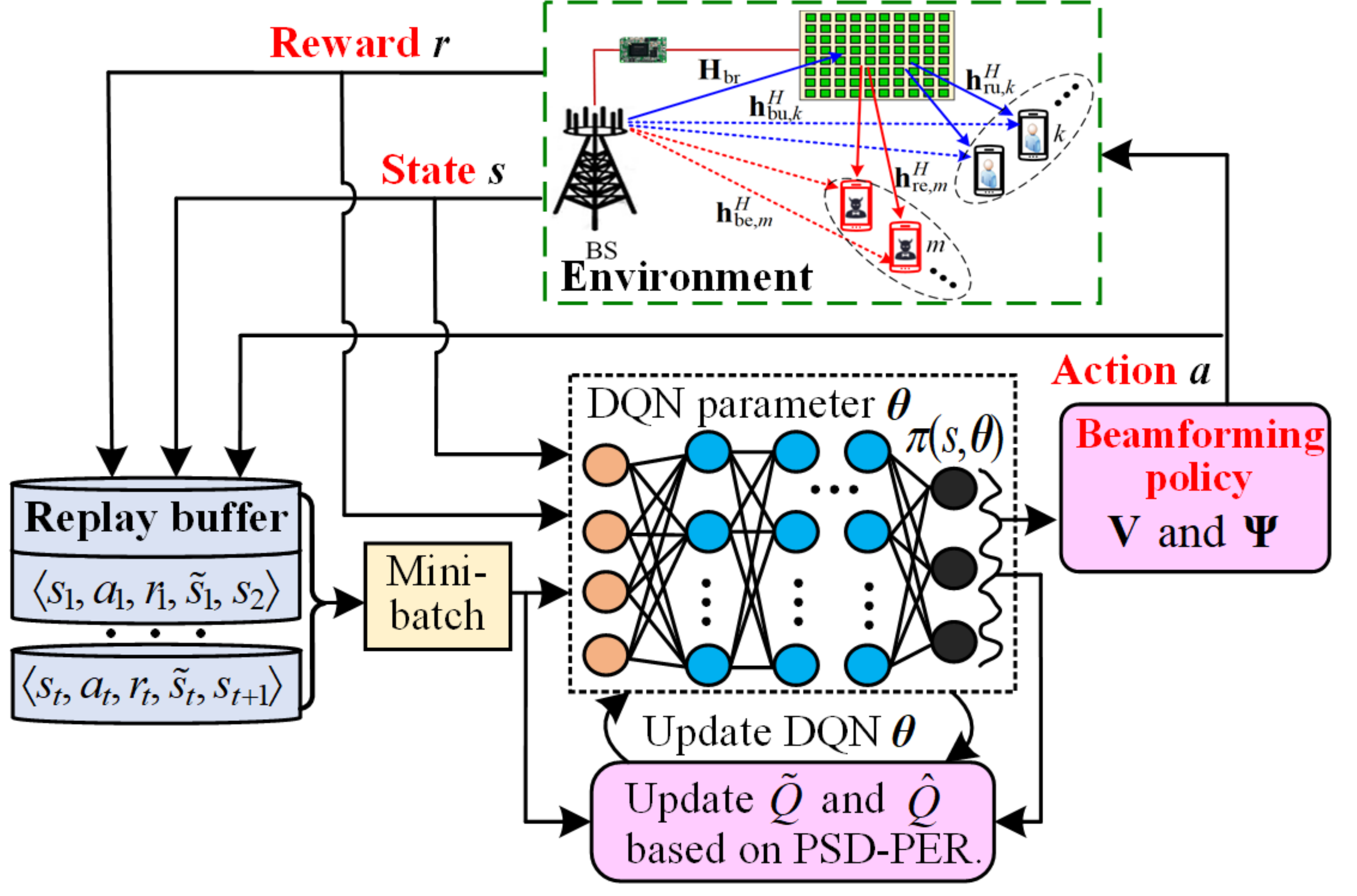}
\caption{{Deep PDS-PER learning based beamforming for IRS-aided
secure communications.} } \label{fig:Schematic}
\end{figure}

Note that the policy optimization (in terms of the BS's
beamforming matrix ${\bf{V}}$ and the RIS's reflecting beamforming
matrix ${\bf{\Psi }}$) in the IRS-aided secure communication
system can be performed at the BS and that the optimized
reflecting beamforming matrix can be transferred in an offline manner to the IRS by the
controller to adjust the corresponding reflecting elements
accordingly.

\subsection{Proposed Deep PDS-PER Learning}

As discussed in Section II, CSI is unlikely to be known accurately
due to the transmission delay, processing delay, and
mobility of users. At the same time, beamforming with outdated CSI
will decrease the secrecy capacity, and therefore, a fast optimization
solution needs to be designed to reduce processing delay.
PDS-learning as a well-known algorithm has been used to improve
the learning speed by exploiting extra partial information (e.g.,
the previous location information and the mobility velocity of MUs
or eavesdroppers that affect the channel coefficients) and search for an optimized policy in dynamic
environments [33]-[35]. Motivated by this, we devise a modified
deep PDS-learning to trace the environment dynamic characteristics, and then adjust the transmit beamforming at the BS and the reflecting elements at the IRS accordingly, which can speed up  the learning efficiency in dynamic
environments.

PDS-learning can be defined as an immediate system state ${\tilde
s_t} \in {\mathcal{S}}$ happens after executing an action
${a_t}$ at the current state ${s_{t}}$  and before the next time state ${s_{t + 1}}$ .
In detail, the PDS-learning agent takes an action  ${a_t}$  at
state ${s_t}$, and then will receive known reward
${r^{\rm{k}}}({s_t},{a_t})$ from the environment before
transitioning the current state ${s_t}$ to the PDS state
${\tilde s_t}$ with a known transition probability
${{\mathcal{T}}^{\rm{k}}}({\tilde s_t}|{s_t},{a_t})$. After that,
the PDS state further transform to the next state ${s_{t + 1}}$
with an unknown transition probability  ${{\mathcal{T}}^{\rm{u}}}({s_{t +
1}}|{\tilde s_t},{a_t})$ and an unknown reward
${r^{\rm{u}}}({s_t},{a_t})$,  which corresponds to the wireless CSI
dynamics.  In PDS-learning, ${s_{t + 1}}$ is independent of
${s_t}$ given the PDS state ${\tilde s_t}$, and the reward
$r({s_t},{a_t})$ is decomposed into the sum of
${r^{\rm{k}}}({s_t},{a_t})$ and ${r^{\rm{u}}}({s_t},{a_t})$ at
${\tilde s_t}$ and ${s_{t + 1}}$, respectively. Mathematically,
the state transition probability in PDS-learning from  ${s_t}$ to
${s_{t + 1}}$ admits
\begin{equation}
\begin{split}
{\mathcal{T}}({s_{t + 1}}|{s_t},{a_t}) = \sum\nolimits_{{{\tilde
s}_t}} {{{\mathcal{T}}^{\rm{u}}}({s_{t + 1}}|{{\tilde
s}_t},{a_t}){{\mathcal{T}}^{\rm{k}}}({{\tilde s}_t}|{s_t},{a_t})}.
\end{split}
\end{equation}
Moreover, it can be verified that the reward of the current
state-action pair $({s_t},{a_t})$ is expressed by
\begin{equation}
\begin{split}
r({s_t},{a_t}) = {r^{\rm{k}}}({s_t},{a_t}) +
\sum\nolimits_{{{\tilde s}_t}} {{{\mathcal{T}}^{\rm{k}}}({{\tilde
s}_t}|{s_t},{a_t}){r^{\rm{u}}}({{\tilde s}_t},{a_t})}.
\end{split}
\end{equation}

At the time slot $t$, the PDS action-value function $\tilde
Q({\tilde s_t},{a_t})$ of the current PDS state-action pair
$({\tilde s_t},{a_t})$ is defined as
\begin{equation}
\begin{split}
\tilde Q({\tilde s_t},{a_t}) = {r^{\rm{u}}}({\tilde s_t},{a_t}) +
\gamma \sum\limits_{{s_{t + 1}}} {{{\mathcal{T}}^{\rm{u}}}({s_{t +
1}}|{{\tilde s}_t},{a_t})V({s_{t + 1}})}.
\end{split}
\end{equation}

By employing the extra information (the known transition
probability ${{\mathcal{T}}^{\rm{k}}}({\tilde s_t}|{s_t},{a_t})$
and known reward   ${r^{\rm{k}}}({s_t},{a_t})$), the Q-function
$\hat Q({s_t},{a_t})$ in PDS-learning can be further expanded
under all state-action pairs $(s,a)$, which is expressed by
\begin{equation}
\begin{split}
\hat Q({s_t},{a_t}) = {r^{\rm{k}}}({s_t},{a_t}) +
\sum\nolimits_{{{\tilde s}_t}} {{{\mathcal{T}}^{\rm{k}}}({{\tilde
s}_t}|{s_t},{a_t})\tilde Q({{\tilde s}_t},{a_t})}.
\end{split}
\end{equation}

 The state-value function in PDS-learning is defined by
\begin{equation}
\begin{split}
{\hat V_t}({s_t}) = \sum\nolimits_{{s_{t + 1}}}
{{{\mathcal{T}}^{\rm{k}}}({s_{t + 1}}|{s_t},{a_t})\tilde V({s_{t +
1}})}
\end{split}
\end{equation}
where  ${\tilde V_t}({s_{t + 1}}) = \mathop {\max }\limits_{{a_t}
\in {\mathcal{A}}} {\tilde Q_t}({\tilde s_{t + 1}},{a_t})$. At
each time slot, the PDS action-value function  $\tilde Q({\tilde
s_t},{a_t})$ is updated by
\begin{equation}
\begin{split}
\begin{array}{l}
{{\tilde Q}_{_{t + 1}}}({{\tilde s}_t},{a_t}) = (1 - {\alpha _t}){{\tilde Q}_t}({{\tilde s}_t},{a_t})\\
\;\;\;\;\;\;\;\;\;\;\;\;\; + {\alpha _t}\left( {{r^{\rm{u}}}({{\tilde s}_t},{a_t}) + \gamma {{\hat V}_t}({s_{t + 1}})} \right)
\end{array}
\end{split}
\end{equation}

After updating  ${\tilde Q_{_{t + 1}}}({\tilde s_t},{a_t})$, the
action-value function  ${\hat Q_{_{t + 1}}}({s_t},{a_t})$ can be
updated by plugging ${\tilde Q_{_{t + 1}}}({\tilde s_t},{a_t})$
into (27).

After presenting in the above modified PDS-learning, a deep PDS
learning algorithm is presented. In the presented learning
algorithm, the traditional DQN is adopted to estimatete the
action-value Q-function $Q(s,a)$ by using  $Q(s,a;{\bm{\theta
}})$, where ${\bm{\theta }}$ denote the  DNN parameter. The objective
of DQN is to minimize the following loss function at each time slot
\begin{equation}
\begin{split}
\begin{array}{l}
{\mathcal{L}}({{\bm{\theta }}_t}) = \left[ {{{\{ {{\hat V}_t}({s_t};{{\bm{\theta }}_t}) - \hat Q({s_t},{a_t};{{\bm{\theta }}_t})\} }^2}} \right] = \left[ {\{ r({s_t},{a_t})} \right.\\
\;\;\;\;\;\;\;\left. { + \gamma \mathop {\max }\limits_{{a_{t + 1}} \in {\mathcal{A}}} {{\hat Q}_t}({s_{t + 1}},{a_{t + 1}};{{\bm{\theta }}_t}) - \hat Q({s_t},{a_t};{{\bm{\theta }}_t}){\} ^2}} \right]
\end{array}
\end{split}
\end{equation}

where ${\hat V_t}({s_t};{{\bm{\theta }}_t}) = r({s_t},{a_t}) +
\gamma \mathop {\max }\limits_{{a_{t + 1}} \in {\mathcal{A}}}
{\hat Q_t}({s_{t + 1}},{a_{t + 1}};{{\bm{\theta }}_t})$ is the
target value. The error between  ${\hat V_t}({s_t};{{\bm{\theta
}}_t})$ and the estimated value  $\hat Q({s_t},{a_t};{{\bm{\theta
}}_t})$ is usually called temporal-difference (TD) error, which is
expressed by
\begin{equation}
\begin{split}
{\delta _t} = {\hat V_t}({s_t};{{\bm{\theta }}_t}) - \hat
Q({s_t},{a_t};{{\bm{\theta }}_t}).
\end{split}
\end{equation}

The DNN parameter  ${\bm{\theta }}$ is achieved by taking the
partial differentiation of the objective function (30) with
respect to ${\bm{\theta }}$, which is given by
\begin{equation}
\begin{split}
{{\bm{\theta }}_{t + 1}} = {{\bm{\theta }}_t} + \beta \nabla
{\mathcal{L}}({{\bm{\theta }}_t}).
\end{split}
\end{equation}
where  $\beta $ is the learning rate of
${\bm{\theta }}$, and $\nabla ( \cdot )$ denotes the first-order
partial derivative.

 Accordingly, the policy ${\hat \pi _t}(s)$ of the modified deep PDS-learning algorithm is given by
\begin{equation}
\begin{split}
{\hat \pi _t}(s) = \arg \mathop {\max }\limits_{{a_t} \in
{\mathcal{A}}} \hat Q({s_t},{a_t};{{\bm{\theta }}_t}).
\end{split}
\end{equation}

Although DQN is capable of performing well in policy learning with
continuous and high-dimensional state space, DNN may learn
ineffectively and cause divergence owing to the nonstationary
targets and correlations between samples. Experience replay is
utilized to avoid the divergence of the RL algorithm. However,
classical DQN uniformly samples each transition ${e_t} = \langle
{s_t},{a_t},{r_t},{\tilde s_t},{s_{t + 1}}\rangle $ from the
experience replay, which may lead to an uncertain or negative
effect on learning a better policy. The reason is that different
transitions (experience information) in the replay buffer have
different importance for the learning policy, and sampling every
transition equally may unavoidably result in inefficient usage of
meaningful transitions. Therefore, a prioritized experience replay
(PER) scheme has been presented to address this issue and enhance
the sampling efficiency [36], [37], where the priority of transition is
determined by the values of TD error. In PER, a transition with
higher absolute TD error has higher priority in the sense that is
has more aggressive correction for the action-value function.

In the deep PDS-PER learning algorithm, similar to classical DQN,
the agent collects and stores each experience  ${e_t} = \langle
{s_t},{a_t},{r_t},{\tilde s_t},{s_{t + 1}}\rangle $ into its
experience replay buffer, and DNN updates the parameter by
sampling a mini-batch of tuples from the replay buffer.  So far,
PER was adopted only for DRL and Q-learning, and has never been
employed with the PDS-learning algorithm to learn the dynamic
information. In this paper, we further extend this PER scheme to
enable prioritized experience replay in the proposed deep PDS-PER
learning framework, in order to improve the learning convergence rate.

The probability of sampling transition $i$ (experience $i$) based
on the absolute TD-error is defined by
\begin{equation}
\begin{split}
p(i) = {{|\delta (i){|^{{\eta _1}}}} \mathord{\left/
 {\vphantom {{|\delta (i){|^{{\eta _1}}}} {\sum\nolimits_{j'} {|\delta (j'){|^{{\eta _1}}}} }}} \right.
 \kern-\nulldelimiterspace} {\sum\nolimits_{j'} {|\delta (j'){|^{{\eta _1}}}} }}
\end{split}
\end{equation}
where the exponent  ${\eta _1}$ weights how much prioritization is
used, with ${\eta _1} = 0$ corresponding to being uniform
sampling. The transition with higher  $p(i)$ will be more likely
to be replayed from the replay buffer, which is associated with
very successful attempts by preventing the DNN from being
over-fitting. With the help of PER, the proposed deep PDS-PER
learning algorithm tends to replay valuable experience and hence
learns more effectively to find the best policy.

It is worth noting that experiences with high absolute TD-error
are more frequently replayed, which alters the visitation
frequency of some experiences and hence causes the training
process of the DNN prone to diverge. To address this problem,
importance-sampling (IS) weights are adopted in the calculation of
weight changes
\begin{equation}
\begin{split}
W(i) = {\left( {D \cdot p(i)} \right)^{ - {\eta _2}}}
\end{split}
\end{equation}
where $D$ is the size of the experience replay buffer, and the
parameter  ${\eta _2}$ is used to adjust the amount of correction
used.

Accordingly, by using the PER scheme into the deep PDS-PER
learning, the DNN loss function (30) and the corresponding parameters are rewritten
respectively as follows:
\begin{equation}
\begin{split}
{\mathcal{L}}\left( {{{\bm{\theta }}_t}} \right) = \frac{1}{H}\sum\limits_{i =
1}^H {\left( {{W_i}{{\mathcal{L}}_i}({{\bm{\theta }}_t})} \right)}
\end{split}
\end{equation}
\begin{equation}
\begin{split}
{{\bm{\theta }}_{t + 1}} = {{\bm{\theta }}_t} + \beta {\delta
_t}{\nabla _{\bm{\theta }}}{\mathcal{L}}({{\bm{\theta }}_t}))
\end{split}
\end{equation}

\emph{Theorem 1:} The presented deep PDS-PER learning can converge
to the optimal  $\hat Q({s_t},{a_t})$ of the MDP with probability
1 when the learning rate sequence ${\alpha _t}$ meets the
following conditions  ${\alpha _t} \in [0,1)$, $\sum\nolimits_{t =
0}^\infty  {{\alpha _t}}  = \infty $ and $\sum\nolimits_{t =
0}^\infty  {\alpha _t^2}  < \infty $, where the aforementioned
requirements have been appeared in most of the RL algorithms [32] and they
are not specific to the proposed deep PDS-PER learning algorithm
[32].

\emph{Proof:} If each action can be executed with an infinite
number of learning steps at each system state, or  in other words, the
learning policy is greedy with the infinite explorations, the
Q-function $\hat Q({s_t},{a_t})$ in PDS-learning and its
corresponding policy strategy $\pi (s)$ will converge
to the optimal points, respectively, with  probability of 1
[33]-[35]. The existing references [34] and [35] have provided the
proof.

\subsection{Secure Beamforming Based on Proposed Deep PDS-PER Learning}

Similar to most DRL algorithms, our proposed deep PDS-PER learning
based secure beamforming approach consists of two stages, i.e., the training stage
and  implementation  stage. The training process of the proposed approach
is shown in \textbf{Algorithm 1}.  A central controller at the BS
is responsible for collecting environment information and making
decision for secure beamforming.

In the training stage, similar to RL-based policy control, the
 controller initializes network parameters and observes the
current system state including CSI of all users, the previous
predicted secrecy rate and the transmission data rate. Then, the
state vector is input into DQN to train the learning model. The
$\varepsilon $-greedy scheme is leveraged to balance both the
exploration and exploitation, i.e., the action with the maximum
reward is selected probability 1- $\varepsilon $  according to the current information (exploitation, which is known knowledge), while a random action is chosen with probability
$\varepsilon $ based on the unknown knowledge (i.e., keep trying new actions, hoping it brings even higher reward (exploration, which is unknown knowledge) ). After executing the
selected action, the agent receives a reward from the environment
and observes the sate transition from  ${s_t}$ to PDS state
${\tilde s_t}$ and then to the next state ${s_{t + 1}}$. Then,
PDS-learning is used to update the PDS action-value function
$\tilde Q({\tilde s_t},{a_t};{{\bm{\theta }}_t})$ and Q-function
$\hat Q({s_t},{a_t};{{\bm{\theta }}_t})$, before collecting and
storing the transition tuple (also called experience)  ${e_t} =
\langle {s_t},{a_t},{r_t},{\tilde s_t},{s_{t + 1}}\rangle $ into
the experience replay memory buffer ${\mathcal{D}}$, which
includes the current system state, selected action, instantaneous
reward and PDS state along with the next state. The experience in
the replay buffer is selected by the PER scheme to generate
mini-batches and they are used to train DQN. In detail, the
priority of each transition $p(i)$ is calculated by using (34) and
then get its IS weight  $W(i)$ in (35), where the priorities
ensure that high-TD-value ($\delta (i)$) transitions are replayed
more frequently. The weight  $W(i)$ is integrated into deep PDS
learning to update both the loss function ${\mathcal{L}}({\bf{\theta }})$ and
DNN parameter  ${\bm{\theta }}$. Once DQN converges, the deep
PDS-PER learning model is achieved.

 After adequate training in \textbf{Algorithm 1}, the learning model is
loaded for the implementation stage. During the implementation stage, the
controller uses the trained learning model to output its selected
action $a$ by going through the DNN parameter  ${\bm{\theta }}$,
with the observed state $s$ from the IRS-aided secure
communication system. Specifically, it chooses an action $a$, with
the maximum value based on the trained deep PDS-PER learning
model. Afterwards, the environment feeds back an instantaneous
reward and a new system state to the agent. Finally, the
beamforming matrix ${{\bf{V}}^ * }$ at the BS and the phase shift
matrix ${{\bf{\Psi }}^ * }$ (reflecting beamforming) at the IRS
are achieved according to the selected action.

We would like to point out  that the training stage needs a powerful
computation server which can be performed offline at the BS while
 the implementation stage can be completed online. The trained learning
model requires to be updated only when the environment (IRS-aided
secure communication system) has experienced greatly changes,
mainly depending on the environment dynamics and service requirements.

\begin{algorithm}[t]
\begin{small}

\caption{\normalsize Deep PDS-PER Learning Based Secure Beamforming}

1:$~$ \textbf{Input:} IRS-aided secure communication simulator and QoS requirements of all MUs (e.g., minimum secrecy rate and transmission rate).\\
2:$~$ \textbf{Initialize:}   DQN with initial Q-function $Q(s,a;{\bm{\theta }})$, parameters  ${\bm{\theta }}$, learning rate   $\alpha $ and  $\beta $.  \\
3: $~$\textbf{Initialize:} experience replay buffer ${\mathcal{D}}$
with size $D$, and mini-batch size $H$.\\
4: $~$ \textbf{for} each episode =1, 2, \dots, ${N^{{\rm{epi}}}}$
\textbf{do}\\
5: $~~$ Observe an initial system state  $s$;\\
6: $~~$ \textbf{for} each time step $t$=0, 1, 2, \dots, $T$ \textbf{do}\\
7: $~~~~$ Select action based on the $\varepsilon $-greedy policy at
current state  ${s_t}$: choose a random action ${a_t}$ with
probability $\varepsilon $;\\
8: $~~~~$ Otherwise,  ${a_t} = \arg \mathop {\max }\limits_{{a_t} \in {\mathcal{A}}} Q({s_t},{a_t};{{\bf{\theta }}_t})$;\\
9: $~~~~$ Execute action  ${a_t}$, receive an immediate reward
${r^{\rm{k}}}({s_t},{a_t})$  and observe the sate transition
from ${s_t}$  to PDS state ${\tilde s_t}$ and then to the next state ${s_{t + 1}}$;\\
10: $~~~$ Update the reward function $r({s_t},{a_t})$ under PDS-learning using (25);\\
11: $~~~$ Update the PDS action-value function  $\tilde Q({\tilde s_t},{a_t};{{\bm{\theta }}_t})$ using (29); \\
12: $~~~$ Update the Q-function $\hat Q({s_t},{a_t};{{\bm{\theta}}_t})$ using (25);\\
13: $~~~$ Store PDS experience ${e_t} = \langle
{s_t},{a_t},{r_t},{\tilde s_t},{s_{t + 1}}\rangle $ in experience
replay buffer ${\mathcal{D}}$,
if  ${\mathcal{D}}$ is full, remove least used experience from ${\mathcal{D}}$;\\
14: $~~~$ \textbf{for} $i$= 1, 2, \dots, $H$ \textbf{do}  \\
 15: $~~~~~$  Sample transition $i$ with the probability $p(i)$ using (34);\\
16: $~~~~~$  Calculate the absolute TD-error $|\delta (i)|$  in  (31);\\
17: $~~~~~$  Update the corresponding IS weight ${W_i}$  using (35);\\
18: $~~~~~$  Update the priority of transition $i$ based on  $|\delta
(i)|$;\\

19: $~~~$ \textbf{end for} \\
20: $~~~$ Update the loss function ${\mathcal{L}}\left( {\bm{\theta }}
\right)$ and parameter ${\bm{\theta }}$ of DQN using (36) and
(37), respectively; \\
21: $~~$ \textbf{end for}\\
22: $~$ \textbf{end for} \\
23: \textbf{Output:}  Return the deep PDS-PER learning model.\\

\end{small}
\label{alg_lirnn}
\end{algorithm}

\subsection{Computational Complexity Analysis}

For the training stage, in DNN, let $L$, ${Z_0}$ and ${Z_l}$
denote the training layers, the size of the input layer (which is
proportional to the number of states) and the number of neurons in
the $l$-th layer, respectively. The computational complexity in
each time step for the agent is  $O({Z_0}{Z_l} + \sum\nolimits_{l
= 1}^{L - 1} {{Z_l}{Z_{l + 1}}} )$. In the training phase, each
mini-batch has ${N^{{\rm{epi}}}}$ episodes with each episode being
$T$ time steps, each trained model is completed iteratively until convergence. Hence, the total computational
complexity in DNN is $O\left( {{N^{{\rm{epi}}}}T({Z_0}{Z_l} +
\sum\nolimits_{l = 1}^{L - 1} {{Z_l}{Z_{l + 1}}} )} \right)$. The
high computational complexity of the DNN training phase can be
performed offline for a finite number of episodes at a centralized
powerful unit (such as the BS).

In our proposed deep PDS-PER learning algorithm, PDS-learning and
PER schemes are utilized to improve the learning efficiency and
enhance the convergence speed, which requires extra computational
complexity. In PDS-learning leaning, since the set of PDS states is
the same as the set of MDP states ${\mathcal{S}}$ [30]-[32], the
computational complexity of the classical DQN algorithm and the
deep PDS-learning algorithm are  $O(|{\mathcal{S}}{|^2} \times
|{\mathcal{A}}|)$ and $O(2|{\mathcal{S}}{|^2} \times
|{\mathcal{A}}|)$, respectively. In PER, since the relay buffer
size is $D$, the system requires to make both updating and
sampling  $O\left( {{{\log }_2}D} \right)$  operations, so the
computational complexity of the PER scheme is  $O\left( {{{\log
}_2}D} \right)$.

According the above analysis, the complexity of the classical DQN
algorithm and the proposed deep PDS-PER learning algorithm are
respectively  $O\left( {I{N^{{\rm{epi}}}}T({Z_0}{Z_l} +
\sum\nolimits_{l = 1}^{L - 1} {{Z_l}{Z_{l + 1}}} ) + |{\mathcal{S}}{|^2}
\times |{\mathcal{A}}|} \right)$ and  $O\left( {I{N^{{\rm{epi}}}}T({Z_0}{Z_l}
+ \sum\nolimits_{l = 1}^{L - 1} {{Z_l}{Z_{l + 1}}} ) + 2|{\mathcal{S}}{|^2}
\times |{\mathcal{A}}| + {{\log }_2}D} \right)$, indicating that the
complexity of the proposed algorithm is slightly higher than the
classical DQN learning algorithm. However, our proposed algorithm
achieves  better performance than  that of  the classical DQN algorithm,
which will be shown in the next section.

\subsection{Implementation Details of DRL }

  This subsection  provides extensive details regarding the generation of training, validation, and testing dataset production.

\textbf{Generation of training:} As shown in Fig. 3, $K$ single-antenna MUs and $M$ single-antenna eavesdroppers are randomly located in the $100 m \times 100 m$  half right-hand side rectangular of Fig. 3 (light blue area) in a two-dimensional x-y rectangular grid plane. The BS and the IRS are
located at (0, 0) and (150, 100) in meter (m), respectively.  The x-y grid has dimensions 100m$\times$100m with a resolution of 2.5 m, i.e., a total of 1600 points.

\begin{figure}
\centering
\includegraphics[width=0.7\columnwidth]{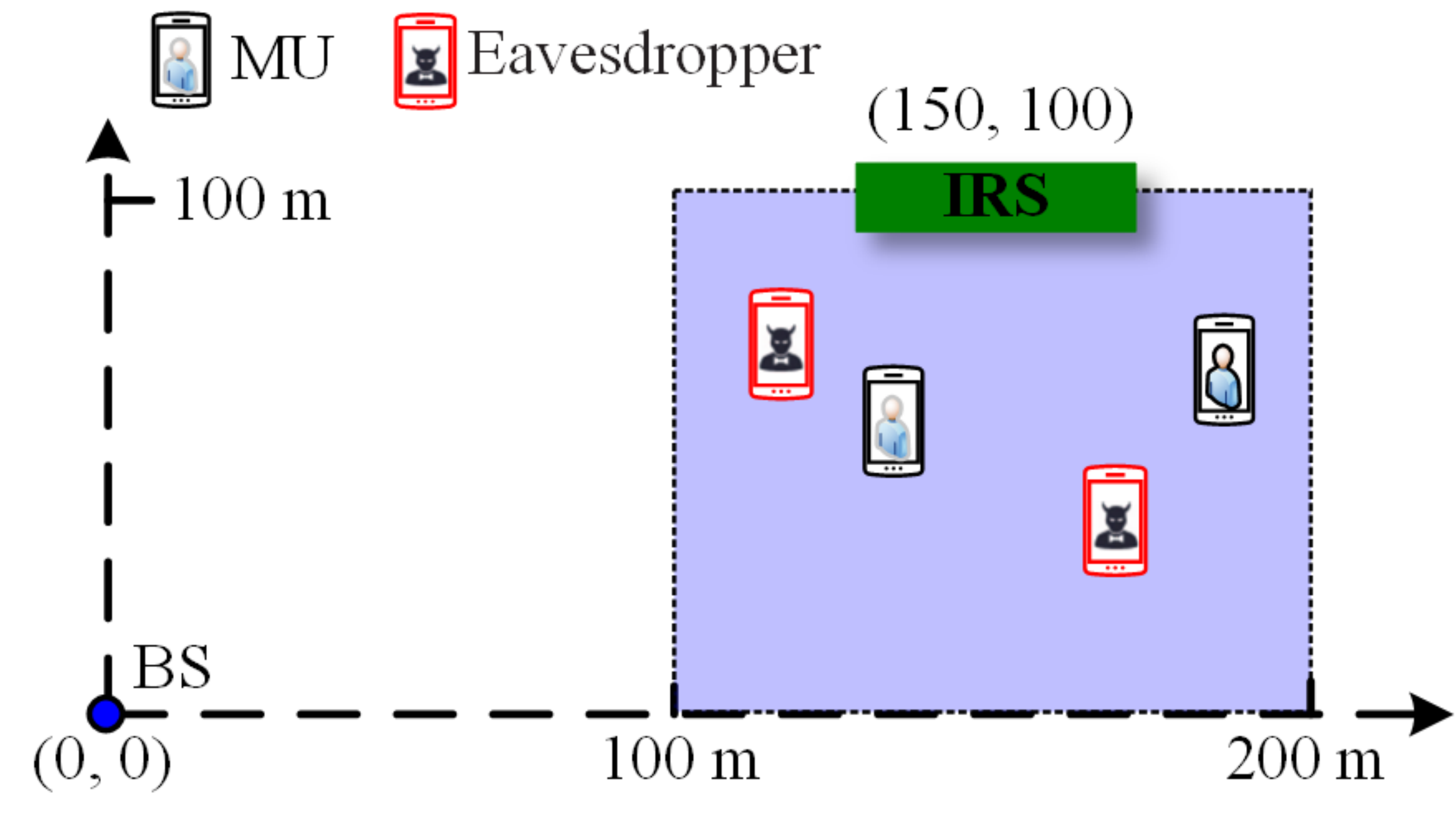}
\caption{{Simulation setup.} } \label{fig:Schematic}
\end{figure}

In the presented IRS-assisted system, the system beamforming codebook ${\mathcal{F}}$  includes the BS's beamforming codebook ${\mathcal{{F_{{\rm{BS}}}}}}$ and the IRS's beamforming codebook ${\mathcal{{F_{{\rm{IRS}}}}}}$. Both the BS's beamforming matrix ${\bf{V}}$ and the IRS's reflection beamforming matrix ${\bf{\Psi }}$ are picked from the pre-defined codebook ${\mathcal{{F_{{\rm{BS}}}}}}$ and  ${\mathcal{{F_{{\rm{IRS}}}}}}$, respectively. The data points of sampled channel vector and the corresponding reward vector $\left\langle {{\bf{h}},{\bf{r}}} \right\rangle$ is added into the DNN training data set  ${\mathcal{D}}$. The sampled channel, ${\bf{h}}$, is the input to DQN.  All the simples are normalized by using the normalization scheme to realize a simple per-dataset scaling. After training, the selected BS's beamforming matrix ${\bf{V}}$ and IRS's beamforming matrix  ${\bf{\Psi }}$ with the highest achievable reward are used to reflect the security communication performance.

The DQN learning model is trained using empirically hyper-parameter, where DNN trained for 1000 epochs with 128 mini-batches being utilized in each epoch. In the training process, 80\% and 20\% of all generated data are selected as the training and validation (test) datasets, respectively. The experience replay buffer size is 32000 where the corresponding samples are randomly sampled from this number of the most recently experiences.

\textbf{DQN structure:} The DQN model is designed as a Multi-Layer Perceptron network, which is also referred to as the feedforward Fully Connected network. Note here that Multi-Layer Perceptron network is widely used to build an advanced estimator, which fulfills the relation between the environment descriptors and the beamforming matrices (both the BS's beamforming matrix and the IRS's reflecting beamforming matrix).

The DQN model is comprised of $L$ layers, as illustrated in Fig. 2, where the first layer is input layer, the last layer is output layer and the remaining layers are the hidden layers. The $l$-th hidden layer in the network has a stack of  neurons, each of which connects all the outputs of the previous layer. Each unit operates on a single input value outputting another single value. The input of the input layer consist of the systems states, i.e., channel samples, the achievable rate and QoS satisfaction level information in the last time slot, while the output layer outputs the predicted reward values with beamforming matrices in terms of the BS's beamforming matrix and the IRS's reflecting beamforming matrix. The DQN construction is used for training stability. The network parameters will be provided in the next section.

\textbf{Training loss function:}  The objective of DRL model is to find the best beamforming matrices, i.e.,  ${\bf{V}}$ and  ${\bf{\Psi }}$, from the beamforming codebook with the highest achievable reward from the environment. In this case, having the highest achievable reward estimation, the regression loss function is adopted to train the learning model, where DNN is trained to make its output,  ${\bf{\hat r}}$, as close as possible to the desired normalized reward,  ${\bf{\bar r}}$. Formally, the training is driven by minimizing the loss function,  ${\mathcal{L}}({\bm{\theta }})$, defined as
\begin{equation}
\begin{split}
{\mathcal{L}}({\bm{\theta }}) = MSE\left( {{\bf{\hat r}},{\bf{\bar r}}} \right),
\end{split}
\end{equation}
where ${\bm{\theta }}$ is the set of all DNN parameters, $MSE(\cdot)$ denotes the mean-squared-error between  ${\bf{\hat r}}$  and ${\bf{\bar r}}$. Note that the outputs of DNN, ${\bf{\hat r}}$ can be acted as functions of ${\bm{\theta }}$ and the inputs of DNN are the system states shown in (12) in the paper.

\section{Simulation Results and Analysis}

This section evaluates the performance of the IRS-aided secure
communication system. The background noise power of MUs and
eavesdroppers is equal to -90 dBm. We set the number of antennas
at the BS is $N=4$, the number of MUs is $K=2$ and the number of
eavesdroppers is $M=2$.  The transmit power  ${P_{\max }}$ at the
BS varies between 15 dBm and 40 dBm, the number of IRS elements
$L$ varies between 10 and 60, and the outdated CSI coefficient
$\rho $ varies from 0.5 to 1 for different simulation settings.
The minimum secrecy rate and the minimum transmission data rate
are 3 bits/s/Hz and 5 bits/s/Hz, respectively. The
path loss model is defined by $PL = \left( {P{L_0} - 10\varsigma
\log 10(d/{d_0})} \right)$  dB, where $P{L_0} = 30\;{\rm{dB}}$
is the path loss at the reference distance  ${d_0} =
1\;{\rm{m}}$ [9], [38],  $\varsigma  = 3$ is the path loss exponent, and
$d$ is the distance from the transmitter to the receiver. The
learning model consists of three connected hidden layers,
containing 500, 250, and 200 neurons [39], respectively. The learning
rate is set to  $\alpha  = 0.002$ and the discount factor is set to
$\gamma  = 0.95$. The the exploration rate $\varepsilon$ is linearly annealed from
0.8 to 0.1 over the beginning 300 episodes and remains
constant afterwards. The parameters ${\mu _1}$ and  ${\mu _2}$ in (12) are set to
${\mu _1} = {\mu _2} = 2$ to balance the utility and cost
[33]-[35]. Similar to the IRS-aided communication  systems [9], [13] and [17], the path loss exponents from the BS to the UEs is set to 3.2, from BS to IRS is set to 2.2, and from IRS to UEs is set to 2.2.

The selection of the network parameters decide the learning convergence speed and efficiency. Here, we take the network parameters, i.e., the learning rate, as an example to demonstrate the importance of the network parameters selection. Fig. 4 shows the average system reward versus training episodes under different learning rates, i.e.,  $\alpha  = \{ 0.1,0.01,0.001,0.0001\}$. It can be observed that different learning rates have different effect on the performance of the deep reinforcement learning algorithm. Specifically, there exits oscillations in behavior for the too-large learning rate of $\alpha  = 0.1$, and its reward performance is much lower than that of $\alpha  = 0.001$. In addition, if we set the too-small learning rate of $\alpha  = 0.0001$, it requires the longer time to achieve the convergence. We can see that the model is able to learn the problem well with the learning rate of  $\alpha  = 0.001$. Hence, we select a suitable learning rate, neither too large nor too small, and the value can be around 0.001 in our test.

\begin{figure}
\centering
\includegraphics[width=0.8\columnwidth]{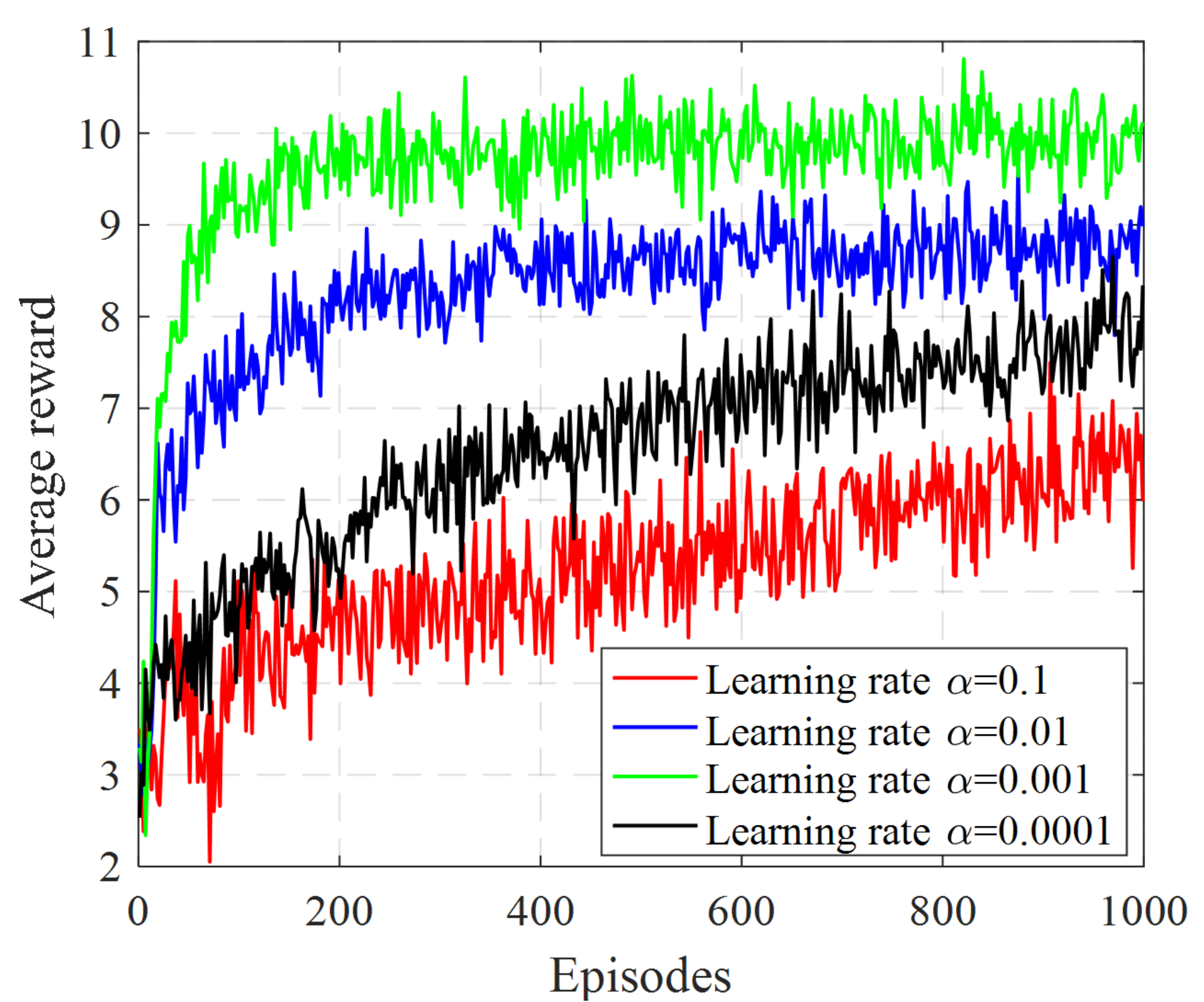}
\caption{{Average reward performance versus episodes under different learning rates.} } \label{fig:Schematic}
\end{figure}

To analyze the loss function of the presented DRL shown in Section IV.D, Fig. 5 illustrates the training/validation loss value during training versus the number of epochs. We can observe that the training/validation loss value decreases significantly in the first few decades epochs and then tend to be approximately at a horizontal level after 150 epochs. Furthermore, the validation loss is only slightly higher than the training loss, which demonstrates that the DNN weights designed have the ability to provide a good fit in terms of the mapping between input samples and output samples. It is worth noting that Fig. 5 is used to investigate how well the DNN weight parameters are designed. If the validation loss is high while the training loss is low, this means that the DQN model is over-fitting and thus the regularization factors may need to be adjusted; if both the validation and training loss values are high, this shows that the DQN model is under-fitting and hence the number of neurons may require to be adjusted.

\begin{figure}
\centering
\includegraphics[width=0.8\columnwidth]{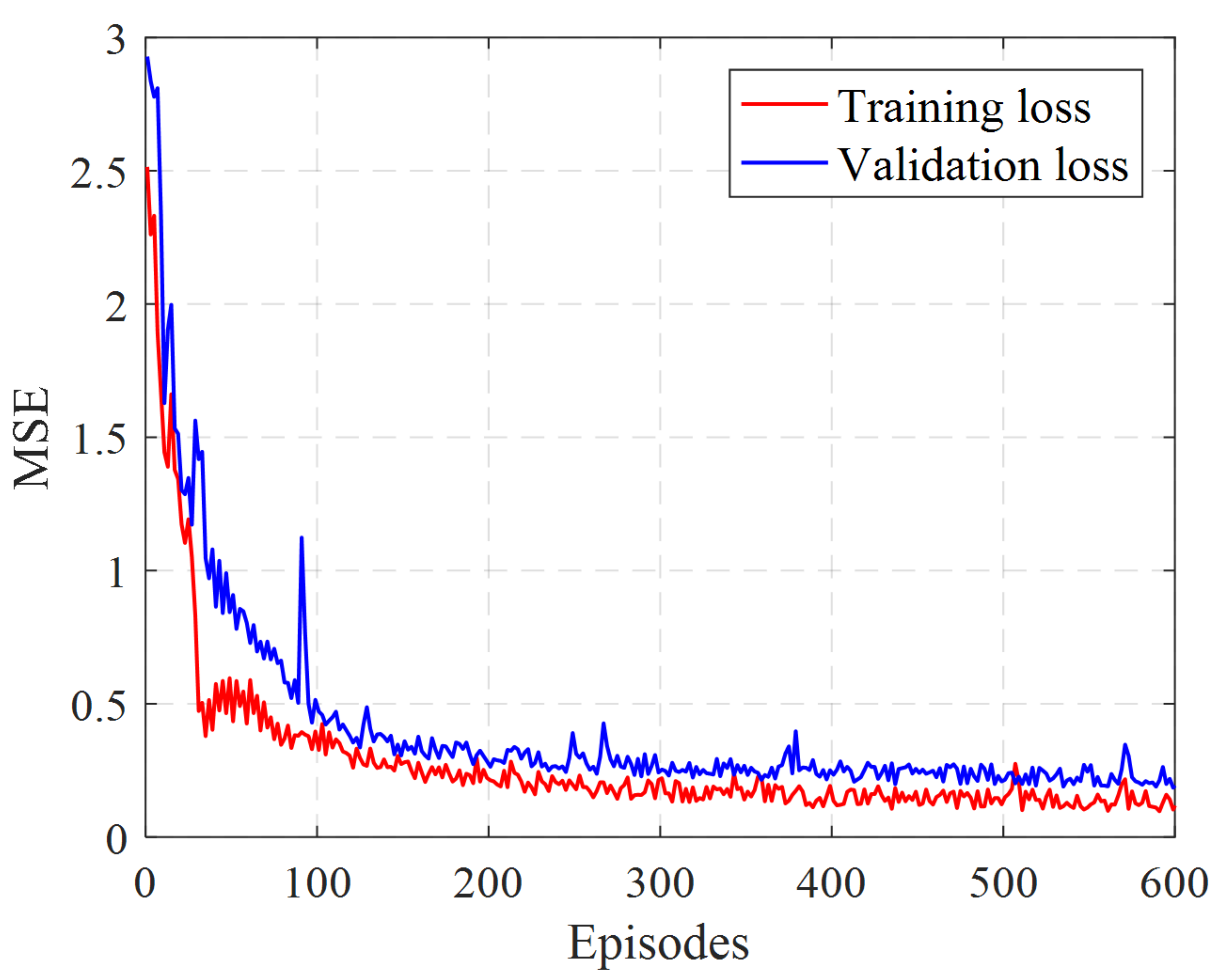}
\caption{{The training and validation losses of DNN employed.} } \label{fig:Schematic}
\end{figure}

In addition, simulation results are
provided to evaluate the performance of the proposed deep PDS-PER learning based
secure beamforming approach (denoted as deep PDS-PER beamforming)
in the IRS-aided secure communication system, and compare the
proposed approach with the following exiting approaches:
\begin{itemize}
\item  The classical DQN based secure
beamforming approach (denoted as DQN-based beamforming), where DNN
is employed to estimate the Q-value function, when acting and
choosing the secure beamforming policy corresponding to the
highest Q-value.
\item   The existing secrecy rate maximization
approach which optimizes the BS's transmit beamforming and the
IRS's reflect beamforming by fixing other parameters as the constants by using the iterative algorithm, which is similar to the suboptimal solution [14] (denoted as Baseline 1 [14]).

\item  The optimal BS's transmit
beamforming approach without IRS assistance (denoted as optimal BS
without IRS). Without IRS, the optimization problem (11) is transformed as
\begin{equation}
\begin{split}
\begin{array}{l}
\mathop {\max }\limits_{{\bf{V}}} \mathop {\min }\limits_{\{ \Delta {\bf{h}}\} } \sum\limits_{k \in {\mathcal{K}}} {R_k^{\sec }} \\
s.t.\;\;({\rm{a}}):\;R_k^{\sec } \ge R_k^{\sec ,\min },\;\forall k \in {\mathcal{K}},\\
\;\;\;\;\;\;\;({\rm{b}}):\;\mathop {\min }\limits_{\scriptstyle||\Delta {{\bf{h}}_{{\rm{bu}}}}|{|^2} \le {({\varsigma _{{\rm{bu}}}})^2},\hfill\atop
\scriptstyle||\Delta {{\bf{h}}_{{\rm{ru}}}}|{|^2} \le {({\varsigma _{{\rm{ru}}}})^2}\hfill} (R_k^{\rm{u}}) \ge R_k^{\min },\;\forall k \in {\mathcal{K}},\\
\;\;\;\;\;\;\;({\rm{c}}):\;{\rm{Tr}}\left( {{\bf{V}}{{\bf{V}}^H}} \right) \le {P_{\max }},\\
\;\;\;\;\;\;\;({\rm{d}}):\;|\chi {e^{j{\theta _l}}}| = 1,\;0 \le {\theta _l} \le 2\pi ,\;\forall l \in {\mathcal{L}}.
\end{array}
\end{split}
\end{equation}

From the optimization problem (39), they system only needs to optimize the BS's transmit beamforming matrix . Problem (39) is non-convex due to the rate constraints, and hence we consider semidefinite programming (SDP) relaxation to solve it. After transforming problem (39), into a convex optimization problem, we can use CVX to obtain the solution [12]-[16].

\end{itemize}

\begin{figure}
\centering
\includegraphics[width=0.75\columnwidth]{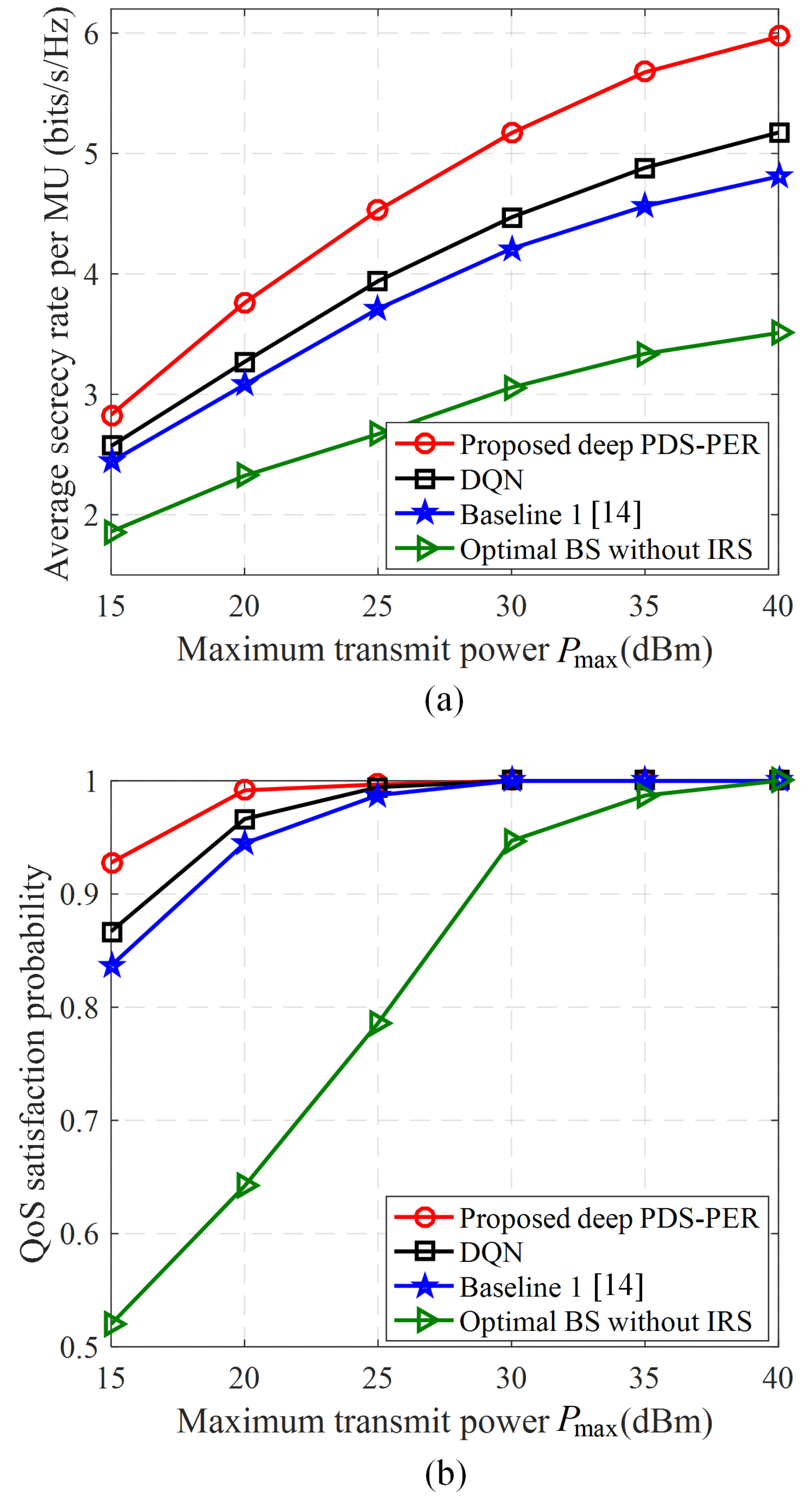}
\caption{{Performance comparisons versus the maximum transmit power
at the BS.} } \label{fig:Schematic}
\end{figure}

Fig. 6 shows the average secrecy rate and QoS satisfaction probability (consists of the secrecy rate satisfaction probability (11a) and the minimum rate satisfaction probability (11b))
versus the maximum transmit power  ${P_{\max }}$, when $L=40$ and
$\rho  = 0.95$.  As expected, both the secrecy rate and QoS
satisfaction probability of all the approaches enhance monotonically
with increasing ${P_{\max }}$. The reason is that when ${P_{\max
}}$ increases, the received SINR at MUs improves, leading to the
performance improvement. In addition, we find that our proposed
learning approach outperforms the Baseline1 approach. In fact, our
approach jointly optimizes the beamforming matrixes ${\bf{V}}$ and
${\bf{\Psi }}$, which can simultaneously facilitates more
favorable channel propagation benefit for MUs and impair
eavesdroppers, while the Baseline1 approach optimizes the
beamforming matrixes in an iterative way. Moreover, our proposed
approach has higher performance than DQN in terms of both secrecy
rate and QoS satisfaction probability, due to its efficient learning
capacity by utilizing PDS-learning and PER schemes in the dynamic
environment. From Fig. 6, we also find that the three IRS assisted
secure beamforming approaches provide significant higher secrecy
rate and QoS satisfaction probability than the traditional system
without IRS. This indicates that the IRS can effectively guarantee
secure communication and QoS requirements via reflecting
beamforming, where reflecting elements (IRS-induced phases) at the
IRS can be adjusted to maximize the received SINR at MUs and
suppress the wiretapped rate at eavesdroppers.

\begin{figure}
\centering
\includegraphics[width=0.75\columnwidth]{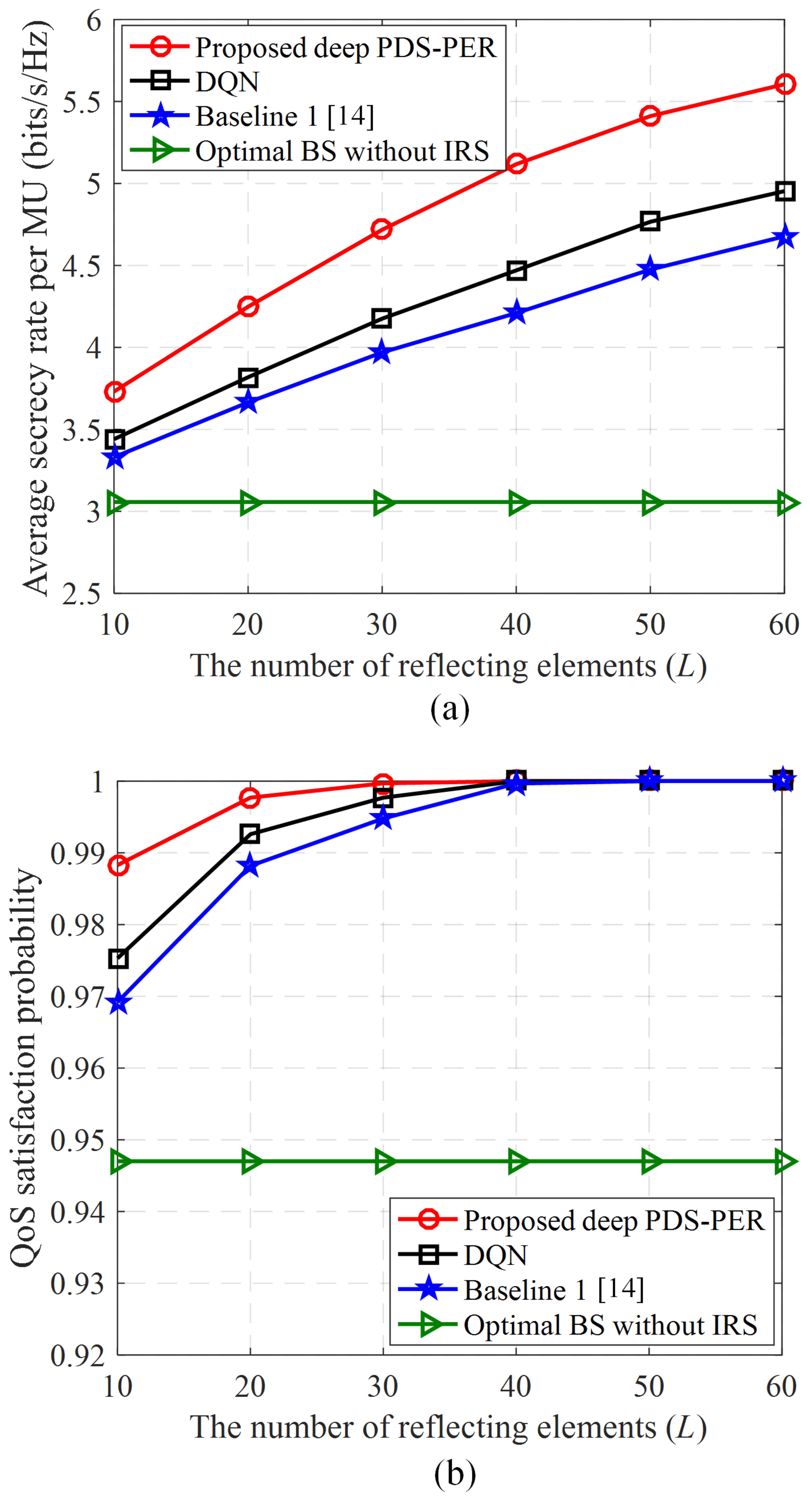}
\caption{{Performance comparisons versus the number of IRS elements.} } \label{fig:Schematic}
\end{figure}

In Fig. 7, the achievable secrecy rate and QoS satisfaction level
performance of all approaches are evaluated through changing the
IRS elements, i.e., from $L=10$ to 60, when ${P_{\max }} =
30\;{\rm{dBm}}$  and $\rho  = 0.95$.   For the secure beamforming
approaches assisted by the IRS, their achievable secrecy rates and
QoS satisfaction levels  significantly increase with the number of
the IRS elements. The improvement results from the fact that more
IRS elements, more signal paths and signal power can be reflected
by the IRS to improve the received SINR at the MUs but to decrease
the received SINR at the eavesdroppers. In addition, the
performance of the approach without IRS remains constant under the
different numbers of the IRS elements.

From Fig. 7(a), it is found that the secrecy rate of the proposed
learning approach is higher than those of the Baseline 1 and DQN
approaches, especially, their performance gap also obviously
increases with $L$, this is because that with more reflecting
elements at the IRS, the proposed deep PDS-PER learning based
secure communication approach becomes more flexible for optimal
phase shift (reflecting beamforming) design and hence achieves
higher gains. In addition, from Fig. 7(b) compared with the
Baseline 1 and DQN approaches, as the reflecting
elements at the IRS increases, we observe that the proposed
learning approach is the first one who attains 100\% QoS
satisfaction level. This superior achievements are based on the
particular design of the QoS-aware reward function shown in (14)
for secure communication.

\begin{figure}
\centering
\includegraphics[width=0.75\columnwidth]{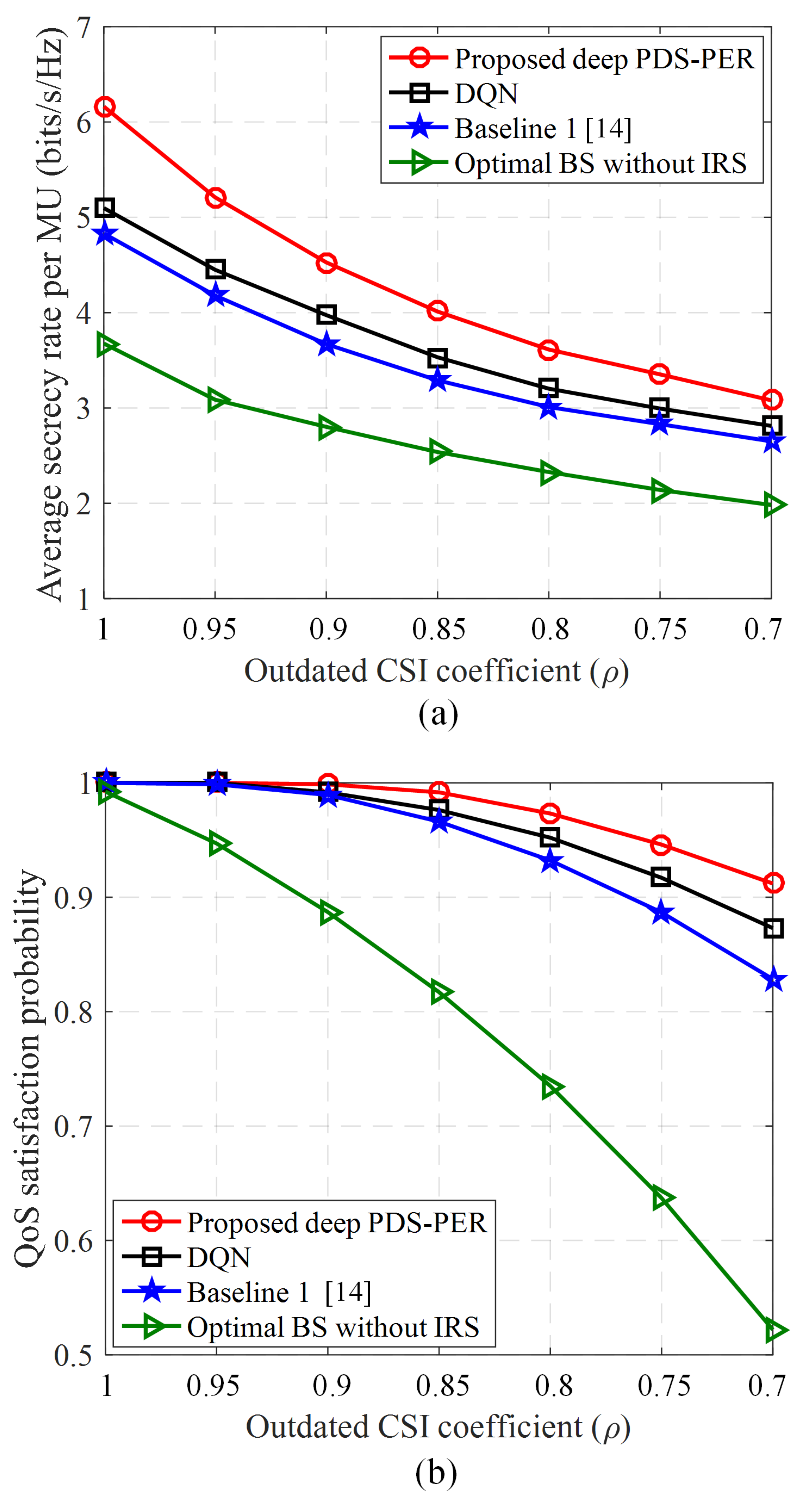}
\caption{{Performance comparisons versus outdated CSI coefficient $\rho$.} } \label{fig:Schematic}
\end{figure}

In Fig. 8, we further analyze how the system secrecy rate and QoS
satisfaction level performances are affected by the outdated CSI
coefficient $\rho$  in the system, i.e., from  $\rho =0.5$ to 1, when
${P_{\max }} = 30\;{\rm{dBm}}$ and $L=40$.  Note that as $\rho$
decreases, the CSI becomes more outdated as shown in (4) and (6),
and  $\rho =1$ means non-outdated CSI. It can be observed from all
beamforming approaches, when CSI becomes more outdated (as $\rho $
decreases), the average secrecy rate and QoS satisfaction level
decrease. The reason is that a higher value of $\rho $ indicates
more accurate CSI, which will enable all the approaches to
optimize secure beamforming policy to achieve higher average
secrecy rate and QoS satisfaction level in the system.

It can be observed that reducing  $\rho $ has more effects on the
performance of the other three approaches while our proposed
learning approach still maintains the performance at a favorable
level, indicating that the other three approaches are more
sensitive to the uncertainty of CSI and the robust of the proposed
learning approach. For instance, the proposed learning approach achieves the secrecy rate and QoS satisfaction level improvements of 17.21\% and 8.67\%, compared with the Baseline 1 approach when $\rho=0.7$. Moreover, in comparison,  the proposed learning
approach achieves the best performance among all approaches against channel uncertainty. The
reason is that the proposed learning approach considers the
time-varying channels and takes advantage of PDS-learning to
effectively learn the dynamic environment.

\section{Conclusion}

In this work, we have investigated the joint BS's beamforming and IRS's
reflect beamforming optimization problem under the time-varying
channel conditions. As the system is highly dynamic and complex, we
have exploited the recent advances of machine learning, and
formulated the secure beamforming optimization problem as an RL
problem. A deep PDS-PER learning based secure beamforming approach
has been proposed to jointly optimize both the BS's beamforming
and the IRS's reflect beamforming in the dynamic IRS-aided secure
communication system, where  PDS and PER schemes have been
utilized to improve the learning convergence rate and efficiency. Simulation
results have verified that the proposed learning approach
outperforms other existing approaches in terms of enhancing the
system secrecy rate and the QoS satisfaction probability.

\begin{IEEEbiography}[{\includegraphics[width=1in,height=1.25in,clip,keepaspectratio]{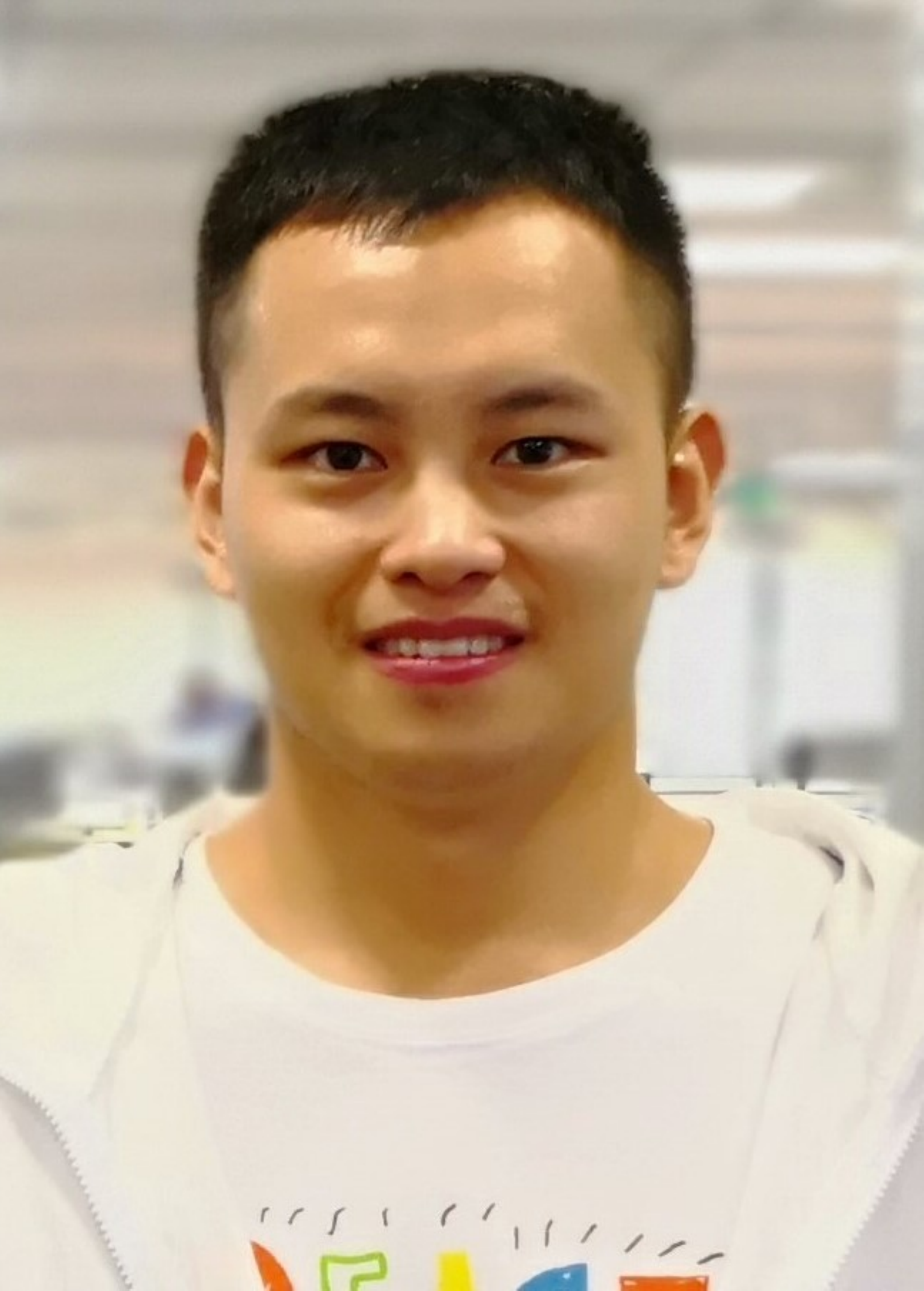}}]{Helin Yang} (S'15) received
 the B.S. and M.S. degrees in the School of
Telecommunications Information Engineering from
Chongqing University of Posts and Telecommunications in 2013, and 2016, and the Ph.D. degree from
School of Electrical and Electronic Engineering,
Nanyang Technological University, Singapore, in
2020. His current research interests include wireless communication, visible light communication, Internet of
Things and resource management.
 \end{IEEEbiography}

\bibliography{bibfile}

\begin{thebibliography}{99}
 

\bibitem{ref:TC01} N. Yang, L. Wang, G. Geraci, M. Elkashlan, J. Yuan, and M. D. Renzo, ``Safeguarding 5G wireless communication networks using physical layer security," \emph{IEEE Commun. Mag.}, vol. 53, no. 4, pp. 20-27, Apr. 2015.
\bibitem{ref:TC01} A. D. Wyner, ``The wiretap channel," \emph{Bell Syst. Tech. J.}, vol. 54, no. 8, pp. 1355--1387, Oct. 1975.
\bibitem{ref:TC01} Q. Li and L. Yang, ``Beamforming for cooperative secure transmission in cognitive two-way relay networks," \emph{IEEE Trans. Inf. Forensics Security}, vol. 15, pp. 130-143, Jan. 2020.
\bibitem{ref:TC01} L. Xiao, X. Lu, D. Xu, Y. Tang, L. Wang, and W. Zhuang, ``UAV relay in VANETs against smart jamming with reinforcement learning," \emph{IEEE Trans. Veh. Technol.}, vol. 67, no. 5, pp. 4087-4097, May 2018.
\bibitem{ref:TC01} W. Wang, K. C. Teh and K. H. Li, ``Artificial noise aided physical layer security in multi-antenna small-cell networks," \emph{IEEE Trans. Inf. Forensics Security}, vol. 12, no. 6, pp. 1470-1482, Jun. 2017.
\bibitem{ref:TC01} H. Wang, T. Zheng, and X. Xia, ``Secure MISO wiretap channels with multiantenna passive eavesdropper: Artificial noise vs. artificial fast fading," \emph{IEEE Trans. Wireless Commun.}, vol. 14, no. 1, pp. 94-106, Jan. 2015.
\bibitem{ref:TC01} R. Nakai and S. Sugiura, ``Physical layer security in buffer-state-based max-ratio relay selection exploiting broadcasting with cooperative beamforming and jamming," \emph{IEEE Trans. Inf. Forensics Security}, vol. 14, no. 2, pp. 431-444, Feb. 2019.
\bibitem{ref:TC01} Z. Mobini, M. Mohammadi, and C. Tellambura, ``Wireless-powered full-duplex relay and friendly jamming for secure cooperative communications," \emph{IEEE Trans. Inf. Forensics Security}, vol. 14, no. 3, pp. 621-634, Mar. 2019.
\bibitem{ref:TC01} Q. Wu and R. Zhang, ``Towards smart and reconfigurable environment: Intelligent reflecting surface aided wireless network," \emph{IEEE Commun. Mag.}, vol. 58, no. 1, pp. 106-112, Jan. 2020.
\bibitem{ref:TC01} J. Zhao, ``A survey of intelligent reflecting surfaces (IRSs): Towards 6G wireless communication networks," 2019. [Online]. Available:  https://arxiv.org/abs/1907.04789.
\bibitem{ref:TC01}  H. Han, \emph{et al.}, ``Intelligent reflecting surface aided power control for physical-layer broadcasting," 2019. [Online]. Available: https://arxiv.org/abs/1912.03468.
\bibitem{ref:TC01}  C. Huang, A. Zappone, G. C. Alexandropoulos, M. Debbah, and C. Yuen, ``Reconfigurable intelligent surfaces for energy efficiency in wireless communication," \emph{IEEE Trans.  Wireless Commun.}, vol. 18, no. 8, pp. 4157-4170, Aug. 2019.
\bibitem{ref:TC01}  Q. Wu and R. Zhang, ``Intelligent reflecting surface enhanced wireless network via joint active and passive beamforming," \emph{IEEE Trans.  Wireless Commun.}, vol. 18, no. 11, pp. 5394-5409, Nov. 2019.


\bibitem{ref:TC01}  M. Cui, G. Zhang, and R. Zhang, ``Secure wireless communication via intelligent reflecting surface," \emph{IEEE Wireless Commun. Lett.}, vol. 8, no. 5, pp. 1410-1414, Oct. 2019.
\bibitem{ref:TC01}  H. Shen, W. Xu, S. Gong, Z. He, and C. Zhao, ``Secrecy rate maximization for intelligent reflecting surface assisted multi-antenna communications," \emph{IEEE Commun. Lett.}, vol. 23, no. 9, pp. 1488-1492, Sep. 2019.
\bibitem{ref:TC01}  X. Yu, D. Xu, and R. Schober, ``Enabling secure wireless communications via intelligent reflecting surfaces," in \emph{Proc. IEEE Global Commun. Conf. (GLOBECOM)}, Waikoloa, HI, USA, Dec. 2019, pp. 1--6
\bibitem{ref:TC01}  Q. Wu and R. Zhang, ``Beamforming optimization for wireless network aided by intelligent reflecting surface with discrete phase shifts," \emph{IEEE Trans. Commun.}, vol. 68, no. 3, pp. 1838-1851, Mar. 2020.
\bibitem{ref:TC01}	Z. Chu, W. Hao, P. Xiao, and J. Shi, ``Intelligent reflecting surface aided multi-antenna secure transmission," \emph{IEEE Wireless Commun. Lett.}, vol. 9, no. 1, pp. 108-112, Jan. 2020.
\bibitem{ref:TC01}	B. Feng, Y. Wu, and M. Zheng, ``Secure transmission strategy for intelligent reflecting surface enhanced wireless system," 2019. [Online]. Available: http://arxiv.org/abs/1909.00629.
\bibitem{ref:TC01}	J. Chen, Y. Liang, Y. Pei and H. Guo, ``Intelligent reflecting surface: A programmable wireless environment for physical layer security," \emph{IEEE Access}, vol. 7, pp. 82599-82612, May 2019.
\bibitem{ref:TC01}	X. Yu, D. Xu, Y. Sun, D. W. K. Ng, and R. Schober, ``Robust and secure wireless communications via intelligent reflecting surfaces," 2019. [Online]. Available: https://arxiv.org/abs/1912.01497.
\bibitem{ref:TC01}	X. Guan, Q. Wu, and R. Zhang, ``Intelligent reflecting surface assisted secrecy communication: Is artificial noise helpful or not?," \emph{IEEE Wireless Commun. Lett.}, vol. 9, no. 6, pp. 778-782, Jun. 2020.

\bibitem{ref:TC01}	L. Dong and H. Wang, ``Secure MIMO transmission via intelligent reflecting surface," Appear in \emph{IEEE Wireless Commun. Lett.}, vol. 9, no. 6, pp. 787-790, June 2020.
\bibitem{ref:TC01}	W. Jiang, Y. Zhang, J. Wu, W. Feng, and Y. Jin, ``Intelligent reflecting surface assisted secure wireless communications with multiple-transmit and multiple-receive antennas," 2019. [Online]. Available: https://arxiv.org/abs/2001.08963.
\bibitem{ref:TC01}	D. Xu, X. Yu, Y. Sun, D. W. K. Ng, and R. Schober, ``Resource allocation for secure IRS-assisted multiuser MISO systems," 2019. [Online]. Available: http://arxiv.org/abs/1907.03085.
\bibitem{ref:TC01}	C. Huang, G. C. Alexandropoulos, C. Yuen, and M. Debbah, ``Indoor signal focusing with deep learning designed reconfigurable intelligent surfaces," 2019. [Online]. Available: https://arxiv.org/abs/1905.07726.
\bibitem{ref:TC01}	A. Taha, M. Alrabeiah, and A. Alkhateeb, ``Enabling large intelligent surfaces with compressive sensing and deep learning," 2019. [Online]. Available: https://arxiv.org/abs/1904.10136.
\bibitem{ref:TC01}	K. Feng, Q. Wang, X. Li and C. Wen, ``Deep reinforcement learning based intelligent reflecting surface optimization for MISO communication systems," \emph{IEEE Wireless Commun. Lett.}, vol. 9, no. 5, pp. 745-749, May 2020.
\bibitem{ref:TC01} C. Huang, R. Mo, and C. Yuen, ``Reconfigurable intelligent surface assisted multiuser MISO systems exploiting deep reinforcement learning," 2020. [Online]. Available: https://arxiv.org/abs/2002.10072.

\bibitem{ref:TC01}	C. Li, W. Zhou, K. Yu, L. Fan, and J. Xia, ``Enhanced secure transmission against intelligent attacks," \emph{IEEE Access}, vol. 7, pp. 53596-53602, Aug. 2019.
\bibitem{ref:TC01}	L. Xiao, G. Sheng, S. Liu, H. Dai, M. Peng, and J. Song, ``Deep reinforcement learning-enabled secure visible light communication against eavesdropping," \emph{IEEE Trans. Commun.}, vol. 67, no. 10, pp. 6994-7005, Oct. 2019.
\bibitem{ref:TC01}	M. Wiering and M. Otterlo, Reinforcement learning: Stateof-the-art, Springer Publishing Company, Incorporated, 2014.
\bibitem{ref:TC01}	H. L. Yang A. Alphones, C. Chen, W. D. Zhong, and X. Z. Xie, ``Learning-based energy-efficient resource management by heterogeneous RF/VLC for ultra-reliable low-latency industrial IoT networks," \emph{IEEE Trans. Ind. Informat.}, vol. 16, no. 8, pp. 5565-5576, Aug. 2020.
\bibitem{ref:TC01}	X. He, R. Jin, and H. Dai, ``Deep PDS-learning for privacy-aware offloading in MEC-enabled IoT," \emph{IEEE Internet of Things J.}, vol. 6, no. 3, pp. 4547-4555, Jun. 2019
\bibitem{ref:TC01}	N. Mastronarde and M. van der Schaar, ``Joint physical-layer and systemlevel power management for delay-sensitive wireless communications," \emph{IEEE Trans. Mobile Comput.}, vol. 12, no. 4, pp. 694-709, Apr. 2013.
\bibitem{ref:TC01}	T. Schaul, J. Quan, I. Antonoglou, and D. Silver,``Prioritized experience replay," in \emph{Proc. 4th Int. Conf. Learn. Represent. (ICLR)}, San Juan, US, May. 2016, pp. 1--21.

\bibitem{ref:TC01}  H. Gacanin and M. Di Renzo, ``Wireless 2.0: Towards an intelligent radio environment empowered by reconfigurable meta-surfaces and artificial intelligence," 2020. [Online]. Available: https://arxiv.org/abs/2002.11040.

\bibitem{ref:TC01} C. W. Huang, $et~al.$, ``Holographic MIMO surfaces for 6G wireless networks: opportunities, challenges, and trends," to appear in \emph{IEEE Wireless Commun.}, Apr. 2020.

\bibitem{ref:TC01}  F. B. Mismar, B. L. Evans, and A. Alkhateeb, ``Deep reinforcement learning for 5G networks: Joint beamforming, power control, and interference coordination,"  \emph{IEEE Trans.  Commun.}, vol. 68, no. 3, pp. 1581-1592, Mar. 2020.

\bibitem{ref:TC01}  H. Yang, $et~al.$, ``Deep reinforcement learning based intelligent reflecting surface for secure wireless communications," will be presented in \emph{Proc. IEEE Global Commun. Conf. (GLOBECOM)}, Dec. 2020, Taipei, Taiwan.



\end{thebibliography}
\begin{IEEEbiography}[{\includegraphics[width=1in,height=1.25in,clip,keepaspectratio]{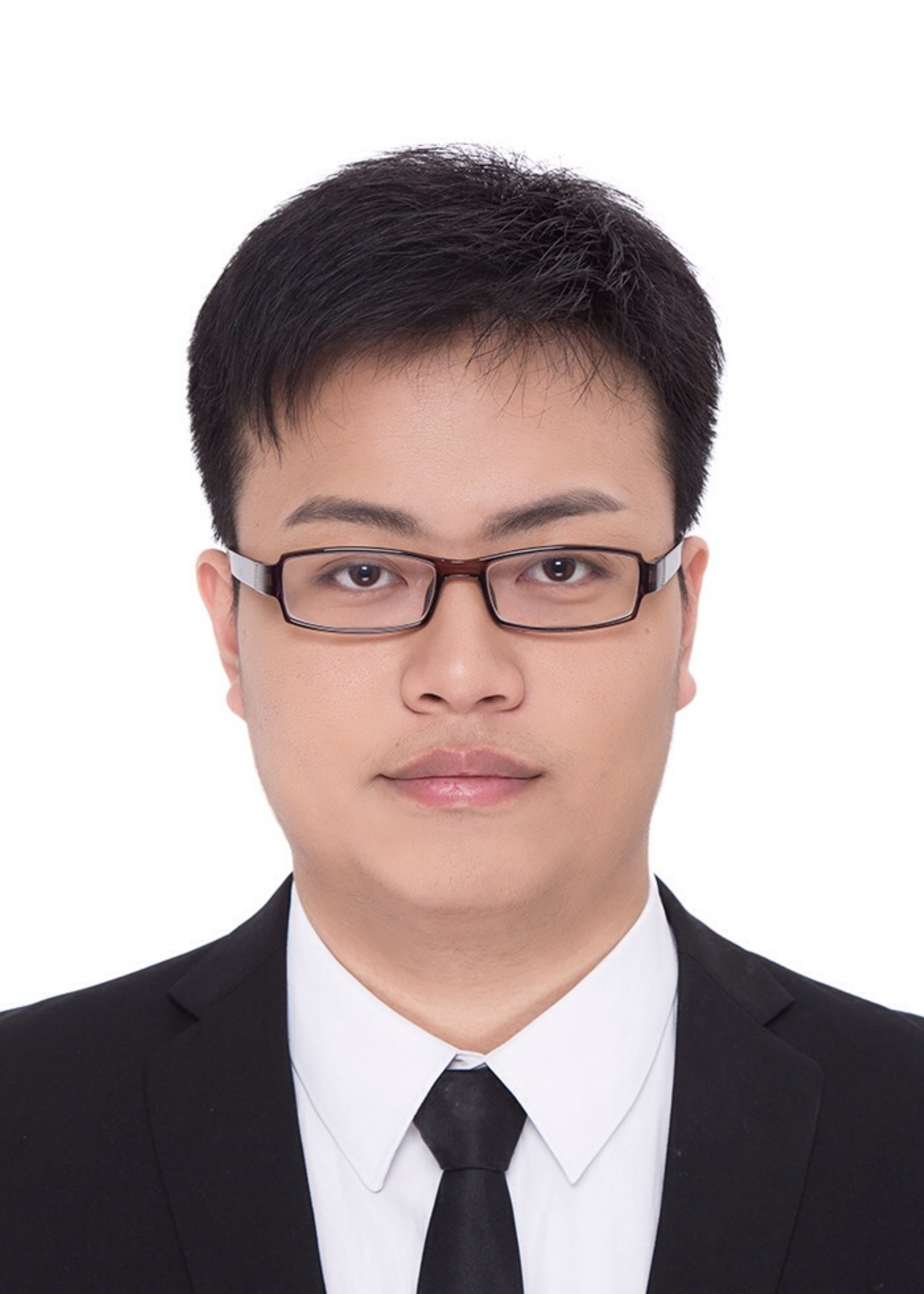}}]{Zehui Xiong}(S'17) is currently a researcher with Alibaba-NTU Singapore Joint Research Institute, Nanyang Technological University, Singapore. He received the Ph.D. degree at Nanyang Technological University, Singapore. He received the B.Eng degree with the highest honors at Huazhong University of Science and Technology, Wuhan, China. He is the visiting scholar with Princeton University and University of Waterloo. His research interests include network economics, wireless communications, blockchain, and edge intelligence. He has published more than 60 peer-reviewed research papers in leading journals and flagship conferences, and 3 of them are ESI Highly Cited Papers. He has won several Best Paper Awards. He is an Editor for Elsevier Computer Networks (COMNET) and Elsevier Physical Communication (PHYCOM), and an Associate Editor for IET Communications. He is the recipient of the Chinese Government Award for Outstanding Students Abroad in 2019, and NTU SCSE Outstanding PhD Thesis Runner-Up Award in 2020.
\end{IEEEbiography}

\begin{IEEEbiography}[{\includegraphics[width=1in,height=1.25in]{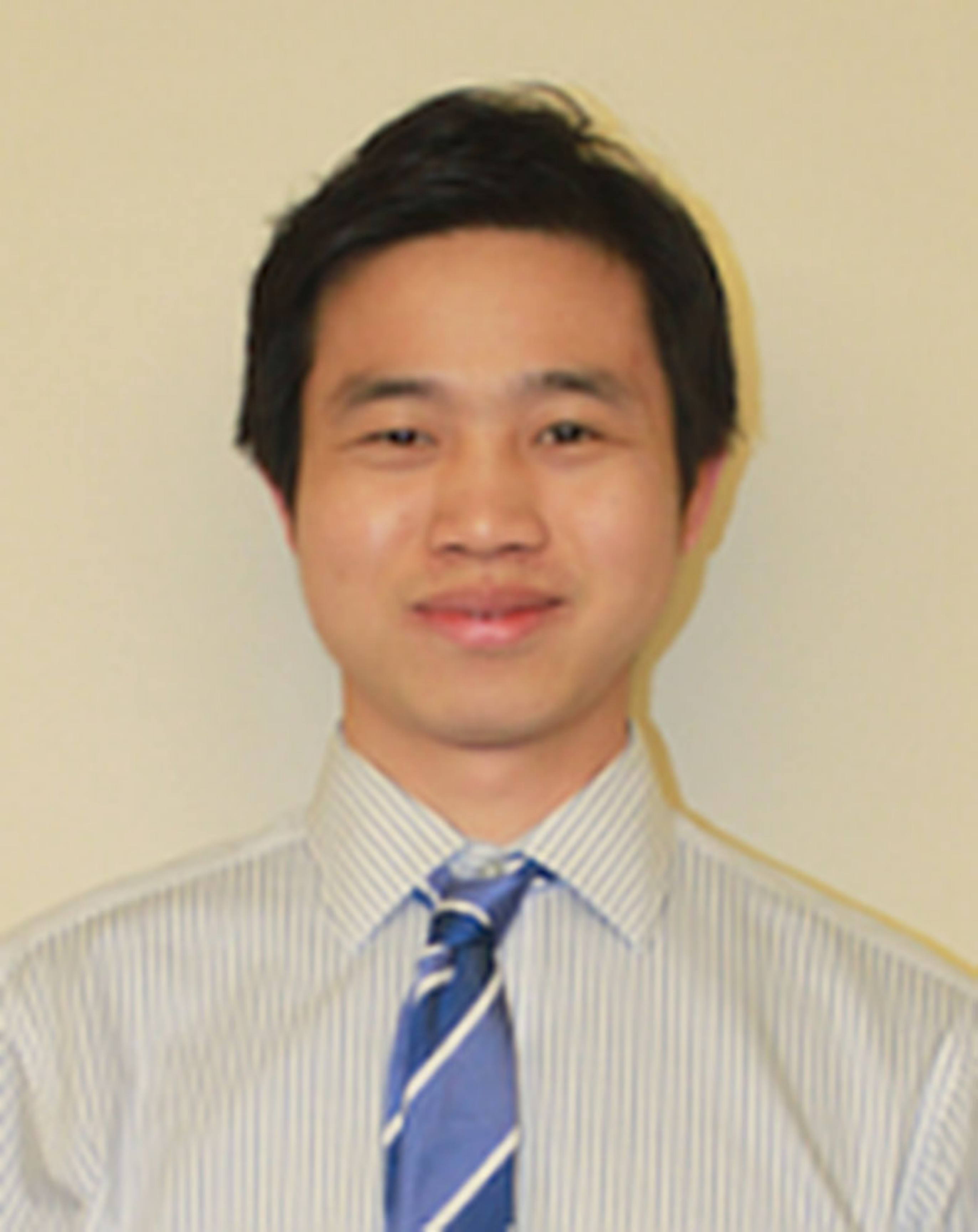}}]{Jun Zhao} (S'10-M'15) is currently an Assistant Professor in the School of Computer Science and Engineering at Nanyang Technological University (NTU) in Singapore. He received a PhD degree in Electrical and Computer Engineering from Carnegie Mellon University (CMU) in the USA (advisors: Virgil Gligor, Osman Yagan; collaborator: Adrian Perrig), affiliating with CMU's renowned CyLab Security \& Privacy Institute, and a bachelor's degree from Shanghai Jiao Tong University in China. Before joining NTU first as a postdoc with Xiaokui Xiao and then as a faculty member, he was a postdoc at Arizona State University as an Arizona Computing PostDoc Best Practices Fellow (advisors: Junshan Zhang, Vincent Poor). His research interests include communications, networks, security, and AI.
\end{IEEEbiography}

\begin{IEEEbiography}[{\includegraphics[width=1in,height=1.25in,clip,keepaspectratio]{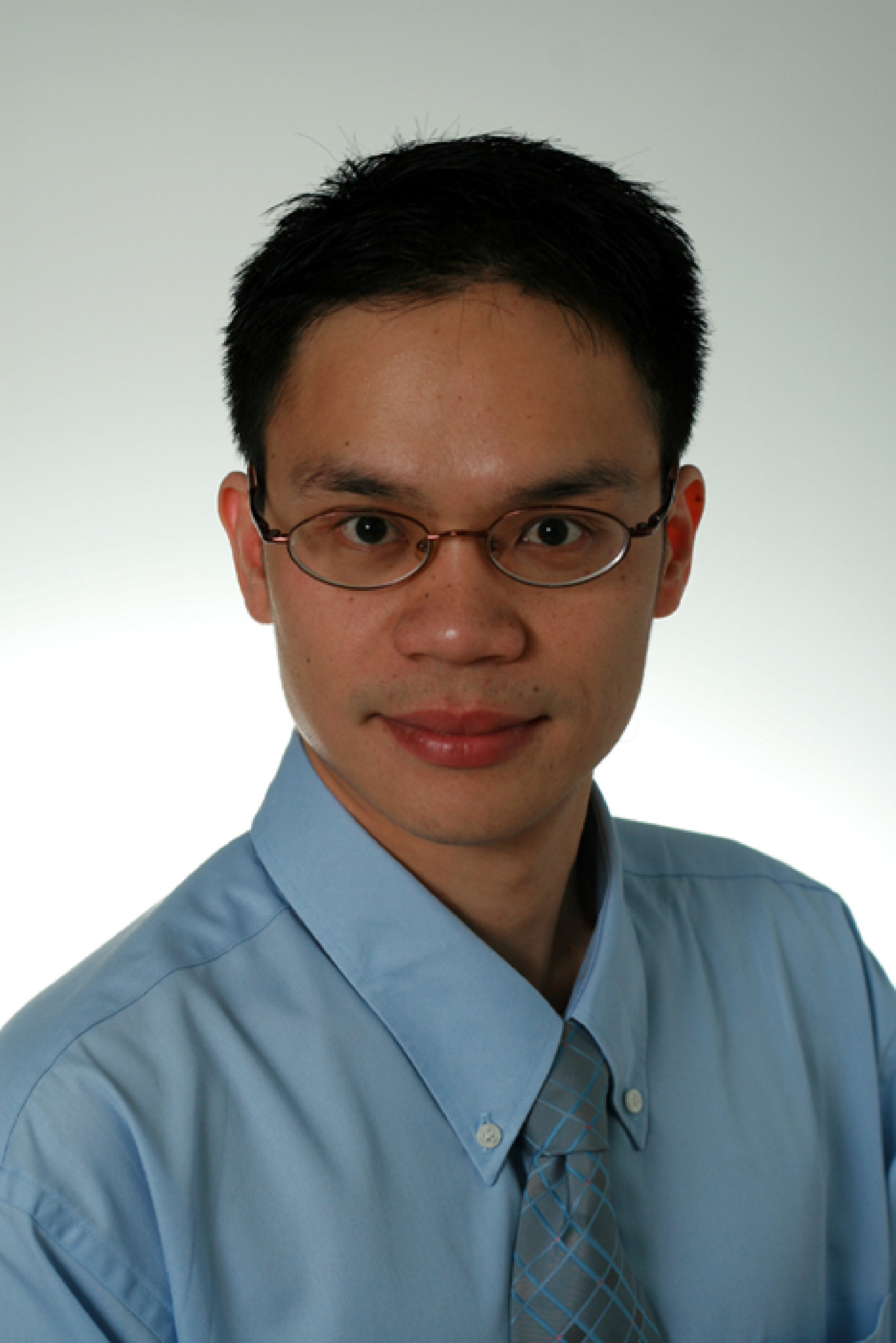}}]{Dusit Niyato} (M'09-SM'15-F'17) is currently a professor in the School of Computer Science and Engineering, at Nanyang Technological University, Singapore. He received B.Eng. from King Mongkuts Institute of Technology Ladkrabang (KMITL), Thailand in 1999 and Ph.D. in Electrical and Computer Engineering from the University of Manitoba, Canada in 2008. His research interests are in the area of energy harvesting for wireless communication, Internet of Things (IoT) and sensor networks.
\end{IEEEbiography}

\begin{IEEEbiography}[{\includegraphics[width=1in,height=1.25in,clip,keepaspectratio]{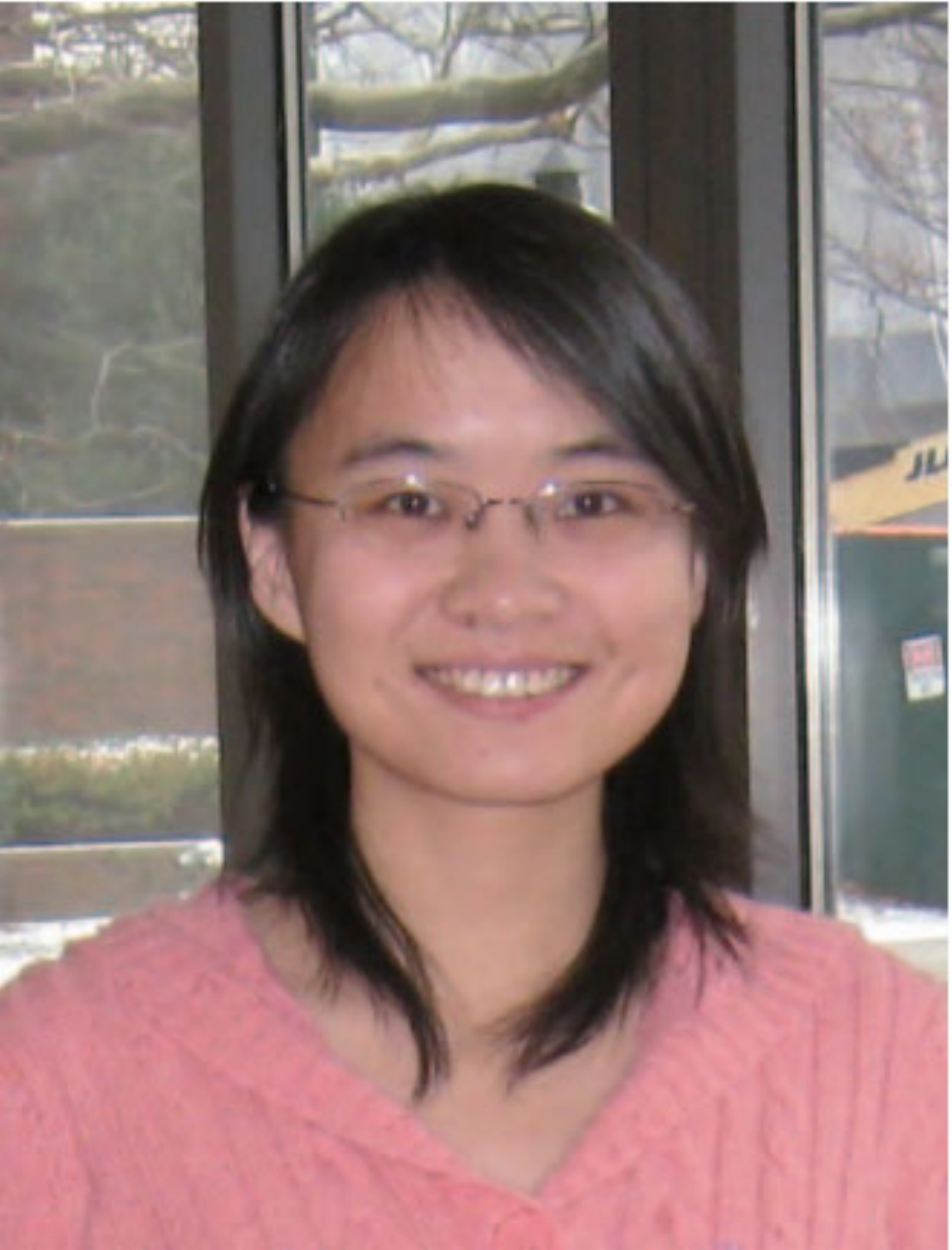}}]{Liang Xiao} (M'09-SM'13) is currently a Professor in the Department of Information and Communication Engineering, Xiamen University, Xiamen, China. She has served as an associate editor of IEEE
Trans. Information Forensics and Security and guest
editor of IEEE Journal of Selected Topics in Signal
Processing. She is the recipient of the best paper award for
2016 INFOCOM Big Security WS and 2017 ICC.
She received the B.S. degree in communication
engineering from Nanjing University of Posts and
Telecommunications, China, in 2000, the M.S. degree in electrical engineering
from Tsinghua University, China, in 2003, and the Ph.D. degree in electrical
engineering from Rutgers University, NJ, in 2009. She was a visiting professor
with Princeton University, Virginia Tech, and University of Maryland, College
Park.

\end{IEEEbiography}

\begin{IEEEbiography}[{\includegraphics[width=1in,height=1.25in,clip,keepaspectratio]{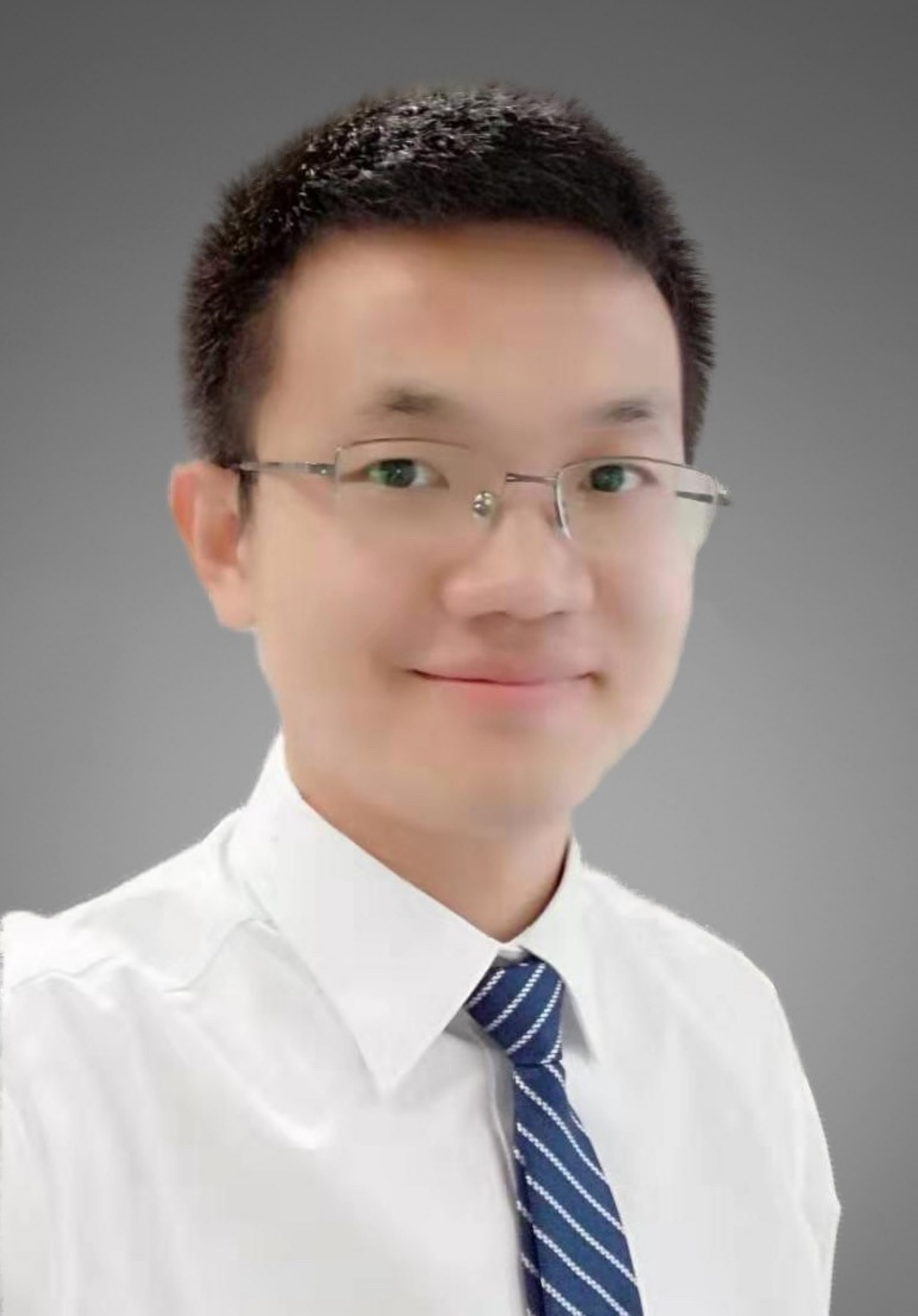}}]{Qingqing Wu} (S'13-M'16) received the B.Eng. and the Ph.D. degrees in Electronic Engineering from South China University of Technology and Shanghai Jiao Tong University (SJTU) in 2012 and 2016, respectively. He is currently an assistant professor in the Department of Electrical and Computer Engineering at the University of Macau, China, and also affiliated with the State key laboratory of Internet of Things for Smart City. He was a Research Fellow in the Department of Electrical and Computer Engineering at National University of Singapore. His current research interest includes intelligent reflecting surface (IRS), unmanned aerial vehicle (UAV) communications, and MIMO transceiver design. He has published over 70 IEEE journal and conference papers. He was the recipient of the IEEE WCSP Best Paper Award in 2015, the Outstanding Ph.D. Thesis Funding in SJTU in 2016, the Outstanding Ph.D. Thesis Award of China Institute of Communications in 2017. He was the Exemplary Editor of IEEE Communications Letters in 2019 and the Exemplary Reviewer of several IEEE journals. He serves as an Associate Editor for IEEE Communications Letters, IEEE Open Journal of Communications Society, and IEEE Open Journal of Vehicular Technology. He is the Lead Guest Editor for IEEE Journal on Selected Areas in Communications on "UAV Communications in 5G and Beyond Networks", and the Guest Editor for IEEE Open Journal on Vehicular Technology on ``6G Intelligent Communications" and IEEE Open Journal of Communications Society on ``Reconfigurable Intelligent Surface-Based Communications for 6G Wireless Networks". He is the workshop co-chair for IEEE ICC 2019-2021 workshop on ``Integrating UAVs into 5G and Beyond", and the workshop co-chair for IEEE GLOBECOM 2020 workshop on ``Reconfigurable Intelligent Surfaces for Wireless Communication for Beyond 5G". He serves as the Workshops and Symposia Officer of Reconfigurable Intelligent Surfaces Emerging Technology Initiative and Research Blog Officer of Aerial Communications Emerging Technology Initiative.
\end{IEEEbiography}

\end{document}